\documentclass[twocolumn, times, twocolSection]{aastex631}

\usepackage{graphicx}
\usepackage{placeins}
\usepackage{color}
\usepackage{tablefootnote}
\usepackage{xcolor}
\usepackage{float}
\usepackage{tabularx}
\usepackage{mhchem}
\usepackage{booktabs} 
\usepackage{soul}

\shorttitle{\texttt{XODIAC}: A Chemical Kinetics Code with Multiple Chemical Networks}
\shortauthors{Ghosh et al.}

\graphicspath{{flowchart/}{benchmark_plots/}{profile_plots/}{performance_plot/}{science_plots/}{chemical_pathways/}{discuss_plots/}{appendix_plot/}}

\begin{document} 

\title{A Fast, Parallelized, GPU-Accelerated Photochemical Model, \texttt{XODIAC}, with Built-in Equilibrium Chemistry and Multiple Chemical Networks for Exoplanetary Atmospheres}

\correspondingauthor{Liton Majumdar}
\email{liton@niser.ac.in; dr.liton.majumdar@gmail.com}

\author[ 0009-0002-4995-9346]{Priyankush Ghosh}
\affiliation{Exoplanets and Planetary Formation Group, School of Earth and Planetary Sciences, National Institute of Science Education and Research, Jatni 752050, Odisha, India}
\affiliation{Homi Bhabha National Institute, Training School Complex, Anushaktinagar, Mumbai 400094, India}

\author[0009-0002-8521-6099]{Sambit Mishra}
\affiliation{Exoplanets and Planetary Formation Group, School of Earth and Planetary Sciences, National Institute of Science Education and Research, Jatni 752050, Odisha, India}
\affiliation{Homi Bhabha National Institute, Training School Complex, Anushaktinagar, Mumbai 400094, India}

\author[0009-0006-0865-8562]{Shubham Dey}
\altaffiliation{These authors contributed equally to this work.}
\affiliation{Exoplanets and Planetary Formation Group, School of Earth and Planetary Sciences, National Institute of Science Education and Research, Jatni 752050, Odisha, India}
\affiliation{Homi Bhabha National Institute, Training School Complex, Anushaktinagar, Mumbai 400094, India}

\author[0009-0001-4100-9218]{Debayan Das}
\altaffiliation{These authors contributed equally to this work.}
\affiliation{Exoplanets and Planetary Formation Group, School of Earth and Planetary Sciences, National Institute of Science Education and Research, Jatni 752050, Odisha, India}
\affiliation{Homi Bhabha National Institute, Training School Complex, Anushaktinagar, Mumbai 400094, India}

\author[0000-0002-7180-081X]{P. B. Rimmer}
\affiliation{Cavendish Laboratory, University of Cambridge, JJ Thomson Ave, Cambridge CB3 0HE, UK}

\author[0000-0001-7031-8039]{Liton Majumdar}
\affiliation{Exoplanets and Planetary Formation Group, School of Earth and Planetary Sciences, National Institute of Science Education and Research, Jatni 752050, Odisha, India}
\affiliation{Homi Bhabha National Institute, Training School Complex, Anushaktinagar, Mumbai 400094, India}

\begin{abstract} 
	The launch of the James Webb Space Telescope (JWST) has delivered high-quality atmospheric observations and expanded the known chemical inventory of exoplanetary atmospheres, opening new avenues for atmospheric chemistry modeling to interpret these data. Here, we present \texttt{XODIAC}, a fast, GPU-accelerated, one-dimensional photochemical model with a built-in equilibrium chemistry solver, an updated thermochemical database, and three chemical reaction networks. This framework enables comparative atmospheric chemistry studies, including the newly developed \texttt{XODIAC-2025} network, a state-of-the-art \texttt{C-H-O-N-P-S-Metals} network, linking 594 species through 7{,}720 reactions. The other two are existing, publicly available \texttt{C-H-O-N-S} and \texttt{C-H-O-N-S-Metals} networks, from the established photochemical models \texttt{VULCAN} and \texttt{ARGO}, respectively, which are commonly used in the community. The \texttt{XODIAC} model has been rigorously benchmarked on the well-studied hot Jupiter HD~189733~b, with results compared against these two models. Benchmarking shows excellent agreement and demonstrates that, when the same chemical network and initial conditions are used, the numerical scheme for solving atmospheric chemistry does not significantly affect the results. We also revisited the atmospheric chemistry of HD~189733~b and performed a comparative analysis across the three networks. Sulfur chemistry shows the least variation across networks, carbon chemistry shows slightly more, and phosphorus chemistry varies the most, primarily due to the introduction of unique \texttt{PHO} and \texttt{PN} pathways comprising 390 reactions in the \texttt{XODIAC-2025} network. These findings highlight \texttt{XODIAC}'s capability to advance exoplanetary atmospheric chemistry and provide a robust framework for comparative exoplanetology.

\end{abstract}


\section{Introduction} \label{sec:intro} 

The discovery of more than 6000 exoplanets (NASA Exoplanet Archive\footnote{\url{https://exoplanetarchive.ipac.caltech.edu/}}) has ushered in a new era of atmospheric characterization, aimed at constraining the physical and chemical processes occurring in their atmospheres through observations of atomic and molecular lines \citep[see, e.g.,][for a review]{Seager10}. However, prior to the \textit{James Webb Space Telescope} (\textit{JWST}), the characterization of exoplanetary atmospheres was mostly limited to gas giants. The chemical inventory of these atmospheres was pioneered primarily by several ground-based and \textit{Hubble Space Telescope} (\textit{HST}) observations using transit spectroscopy \citep[see, e.g.,][for a review]{Madhu19}, which were largely restricted to detecting water (H$_2$O) and alkali species (e.g., Na, K). Within this limited inventory, only H$_2$O was routinely detected using the Wide Field Camera~3 (WFC3) on \textit{HST} in the near-infrared \citep{Sing16}. Secure detections of other key absorbers at longer wavelengths, such as methane (CH$_4$), carbon monoxide (CO), and carbon dioxide (CO$_2$), have remained elusive \citep{Deming17}.

The launch of \textit{JWST} in 2021 marked the beginning of a new chapter in exoplanetary science, providing an abundance of high-quality atmospheric observations and expanding the known chemical inventory more than ever before \citep[see, e.g.,][for a review]{fortney2024characterizing}. Thanks to the broader wavelength coverage and high sensitivity of \textit{JWST}, pioneering discoveries include the detections of CO$_2$ and SO$_2$ in WASP-39~b, indicating complex atmospheric chemistry with photochemical processes \citep{Tsai2023, alderson2023early, rustamkulov2023early}, and CH$_4$ in WASP-80~b, suggesting a solar to sub-solar C/O ratio and roughly five times the solar metallicity in its atmosphere \citep{bell2023methane}. These findings have opened a new era in planetary atmospheric chemistry modeling aimed at interpreting such observations.

To date, one-dimensional (1D) photochemical models, which account for chemical disequilibrium through a chemical kinetics framework incorporating vertical transport of atmospheric gases and photochemical dissociation driven by the host star's irradiation, have served as the primary tool for atmospheric chemistry investigations. These models have been successfully applied to various classes of exoplanets. Among them, several studies on hot Jupiters \citep{Zahnle_2009, Mosses_2011, Venot_2012, Rimmer_2016, Tsai_2017, Hobbs_2019, Tsai_2021} have shown that disequilibrium processes drive gas-phase chemistry away from local chemical equilibrium, leaving clear signatures in either simulated observables or existing observations. 

Although modeling gas-phase atmospheric chemistry using two-dimensional (2D; \citealt{2014A&A...564A..73A, shami24}) and three-dimensional (3D; \citealt{2020A&A...636A..68D, 2023A&A...672A.110L}) transport models has been explored, the need for one-dimensional (1D) photochemical models remains critical because they are simpler, faster, and computationally less demanding. Moreover, these models are particularly well suited to handle large chemical networks with accurately determined rate coefficients and to incorporate disequilibrium processes. This need is expected to grow in importance with the anticipated expansion of the chemical inventory to include more complex molecules and their isotopologues in the era of \textit{Extremely Large Telescopes} (ELTs) \citep{Ignas2025}. Moreover, the chemical inventory for other classes of exoplanets, such as sub-Neptunes and rocky exoplanets, including lava worlds, has already begun to expand with \textit{JWST} observations \citep{Esp2025} and is expected to grow further in the ELT era \citep{Dubye25}. These advances further highlight the need for atmospheric chemistry networks valid across a broad temperature range, from very low to extremely high values (100--30{,}000~K) \citep{Rimmer_2016}.

However, a fundamental challenge for 1D photochemical models arises when the rate constants in the chemical network are valid only over narrow temperature ranges. In such cases, the model may be well-suited to studying specific classes of exoplanets, but not all. Furthermore, if the chemical network is too large, it can suffer from convergence issues and long computational times, making it difficult to obtain accurate results within a reasonable timeframe \citep{Al24}. In addition, all currently available photochemical models in the community lack a standardized format for reaction rate data, resulting in substantial difficulties when transferring chemical networks between different modeling frameworks. Collectively, these limitations have led published 1D photochemical models to rely on their own native chemical networks, such as \texttt{C-H-O} \citep{2003ApJ...596L.247L}, \texttt{C-H-O-N-S} \citep{Zahnle_2009, Hu_2012, Tsai_2021, Hobbs_2021}, \texttt{C-H-O-N} \citep{Mosses_2011, Venot_2012}, and \texttt{C-H-O-N-S-Metals} \citep{Rimmer_2016, Rimmer_2021}, within their respective frameworks. Consequently, it is extremely challenging to apply one network in another model for comparative studies, making it difficult to assess how the choice of chemical network and its associated rate coefficients affect the results. Even for a given chemical network and identical model inputs, differences in numerical schemes, solvers, discretization methods, treatments of vertical diffusion, and convergence criteria can also influence the outcomes.

Addressing these challenges requires a 1D photochemical model that can: (1) run large chemical networks efficiently, (2) employ a stable solver for robust convergence, (3) flexibly adopt different chemical networks, and (4) include a state-of-the-art network covering a broad temperature range. To meet this need, we introduce \texttt{XODIAC} \textit{(eXOplanetary model for equilibrium and DIsequilibrium Atmospheric Chemistry)}, a fully Python-based 1D photochemical kinetics model with a built-in equilibrium chemistry solver, with an updated thermochemical database. \texttt{XODIAC} includes three chemical networks: the \texttt{C-H-O-N-S} network from \citet{Tsai_2021}, the \texttt{C-H-O-N-S-Metals} network from \citet{Rimmer_2021}, and a new state-of-the-art \texttt{C-H-O-N-P-S-Metals} network introduced here. We demonstrate \texttt{XODIAC} by applying it to the well-studied and well-characterized hot Jupiter HD~189733b, which has been extensively investigated both theoretically and observationally \citep{Hobbs_2019, Tsai_2021, fu2024hydrogensulfidemetalenrichedatmosphere}.  

The paper is organized as follows. Section~\ref{sec:meth} describes our model and methodology. Section~\ref{subsec: XODIAC Validation} benchmarks \texttt{XODIAC} against two widely used 1D photochemical codes, \texttt{ARGO} \citep{Rimmer_2016} and \texttt{VULCAN} \citep{Tsai_2021}, and compares their computational performance. We further demonstrate the flexibility of \texttt{XODIAC} when applied to different chemical networks. Section~\ref{sec:res} presents and discusses the results, with particular attention to the behavior of carbon-, sulfur-, and phosphorus-bearing species at different altitudes for each chemical network. Finally, Section~\ref{sec:con} summarizes our conclusions.
% Section 5 depicts an overall discussion of the model and its newly implemented chemistry.  
\begin{figure*}
	\centering
	\includegraphics[width=1\linewidth]{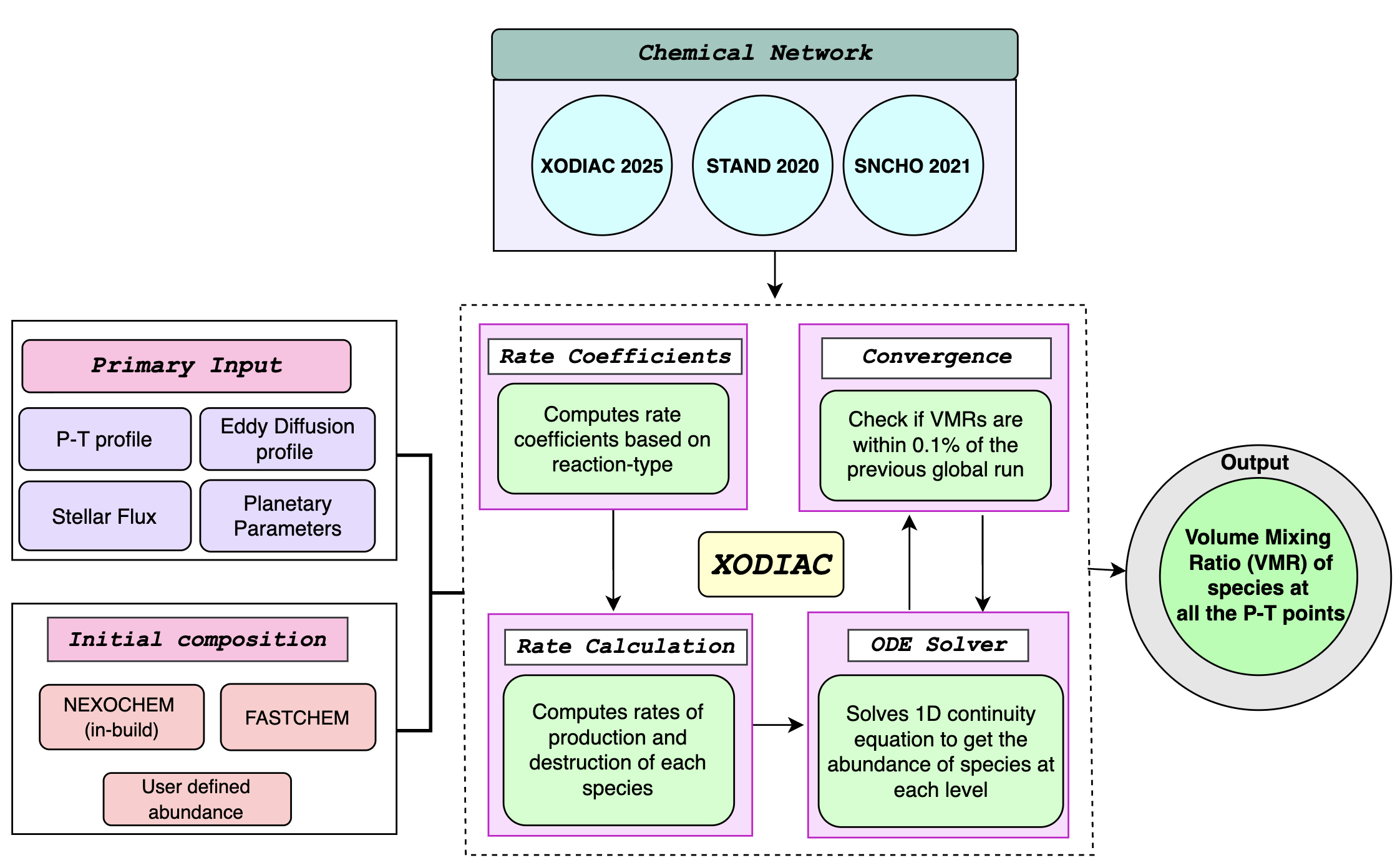}
	\caption{Schematic overview of \texttt{XODIAC}, illustrating the available options for initial composition, chemical networks, pressure–temperature (P–T) profiles, eddy diffusion ($k_{zz}$), actinic flux, and planetary parameters. The model employs a Lagrangian methodology and an ODE solver to compute species volume mixing ratio (VMR) profiles until convergence.}
	\label{fig:xodiac-flowchart}
\end{figure*}

\section{The \texttt{XODIAC} Photochemical Model} \label{sec:meth} 

\begin{figure}
	\centering
	\includegraphics[width=\linewidth]{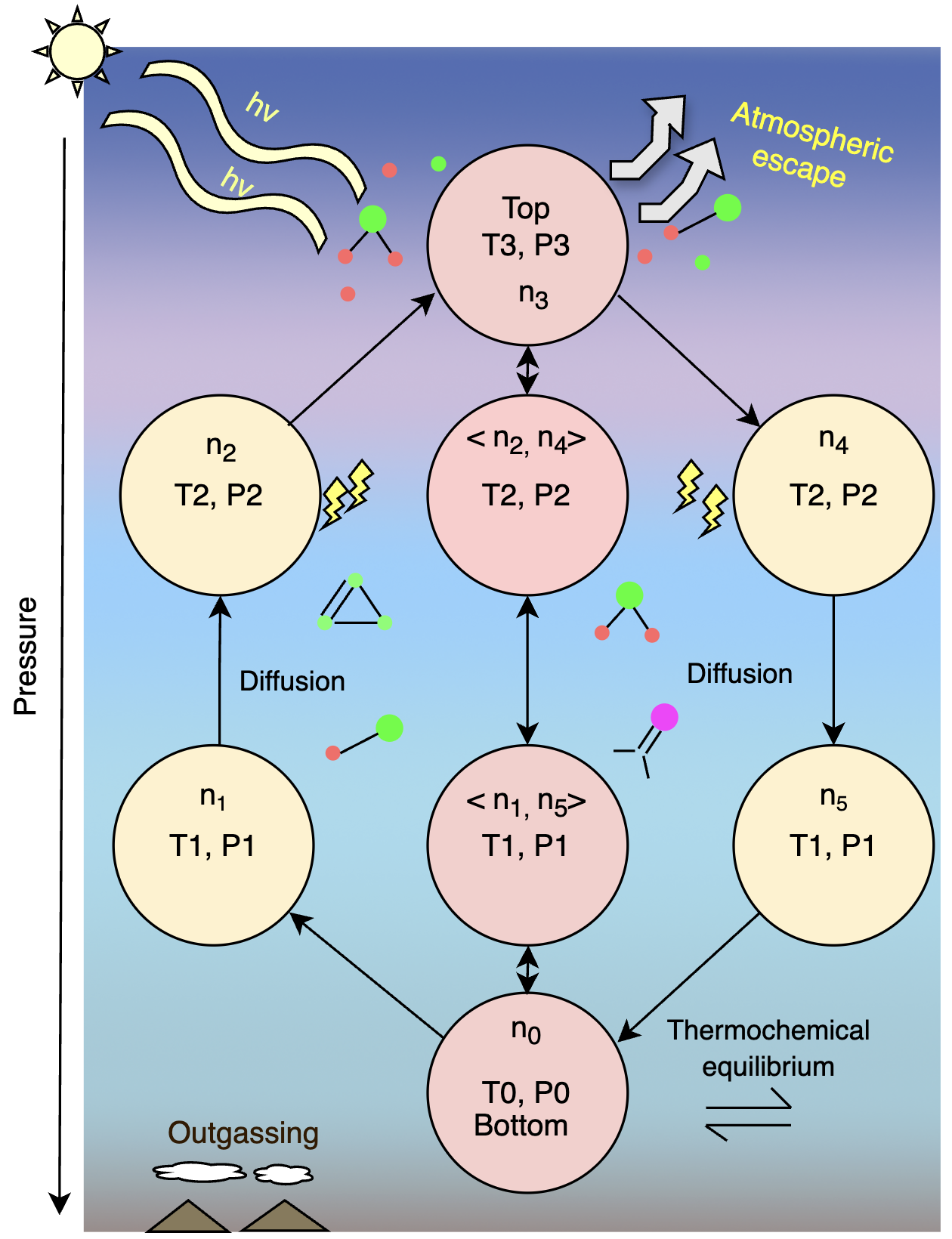}
	\caption{The figure shows the movement of an air parcel in a one-dimensional atmosphere transitioning from the lower atmosphere up to the higher atmosphere with volume mixing ratio (VMR) varying due to molecular diffusion, vertical mixing, photochemistry, and atmospheric escape.} 
	\label{fig: air parcel}
\end{figure}

\texttt{XODIAC} is a fully Python-based photochemical kinetics model that solves the one-dimensional (1-D) continuity equation using a Lagrangian scheme and evolves the atmosphere by simulating various physicochemical processes until a steady-state solution is achieved. The Lagrangian scheme provides an effective framework for accurately tracking the evolution of an air parcel, as it naturally preserves its chemical history and physical conditions along its trajectory, and is particularly well suited for handling large chemical networks. The flowchart illustrating the computational framework of the \texttt{XODIAC} photochemical kinetics model is presented in Figure~\ref{fig:xodiac-flowchart}. In this scheme, during a single global iteration, we track an air parcel from the bottom of the atmosphere (BOA) to the top of the atmosphere (TOA) and back to the BOA \citep{zahnle_et_al...1995, 2008MWRv..136.4653A, Rimmer_2016}. Convergence is assessed following the approach of \citet{Rimmer_2016}, wherein the model is considered converged when the VMR profiles from successive global iterations differ by less than 0.1\% at all altitudes \citep{Rimmer_2016}. A schematic representation of the movement of an air parcel is shown in Figure~\ref{fig: air parcel}. 

\subsection{Numerical Method and the Python-Based ODE Solver \texttt{pylsodes}} \label{subsec: XODIAC Numeric} %\label{subsubsec:continuity equation}
The fundamental equation underlying our model is the one-dimensional (1-D) vertical continuity equation, which describes the time-dependent atmospheric chemistry in the vertical direction. It can be written as  
\vspace{0.3cm}

\begin{equation}
	\frac{\partial n_y}{\partial t} = P_y - L_y - \frac{\partial \phi_y}{\partial z},
	\label{eq:continuity}
\end{equation}
\vspace{0.3cm}

where $y = 1, \ldots, S$ is the index of the $y$th species ($S$ being the total number of species), $n_y$ is its number density, $P_y$ is its production rate, $L_y$ is its loss rate, and $\phi_y$ is the vertical transport flux. This transport flux arises from both eddy diffusion and molecular diffusion. It can be expressed as \citep{1973aero.book.....B, Rimmer_2016, Rimmer_2021}:
\vspace{0.3cm}

\begin{equation}
	\phi_y = -K \left( \frac{\partial n_y}{\partial z} + n_y \left( \frac{1}{H_o} + \frac{1}{T} \frac{dT}{dz} \right) \right)
	\label{eq:transportflux}
\end{equation}
\vspace{0.3 cm}

\hspace{1.7 cm}$- D \left( \frac{\partial n_y}{\partial z} + n_y \left( \frac{1}{H_y} + \frac{(1 + \alpha_T)}{T} \frac{dT}{dz} \right) \right)$
\vspace{0.3 cm}

where $H_o$ is the pressure scale height of the atmosphere, $H_y$ is the molecular scale height of the atmosphere for species $y$, $\alpha_T$ is the thermal diffusion factor, and $z$ is the atmospheric altitude.

The molecular diffusion coefficients are calculated using either Chapman--Enskog theory \citep{enskog1917, chapman1991mathematical} or the theory of binary mixtures, depending on the chosen atmospheric base gas \citep{1972JPCRD...1....3M, Moses2000_icarus, 1973aero.book.....B, Hu_2012, Tsai_2021}. The former requires knowledge of the molecular masses, van der Waals collision cross-sections, thermal diffusivities of the diffusing gases, and the temperature and pressure at each atmospheric layer. The latter requires only the molecular masses, layer temperatures, and layer densities. In the current implementation, the binary mixture formulation allows the base gas to be specified as $\mathrm{H_2}$, $\mathrm{N_2}$, $\mathrm{CO_2}$, or $\mathrm{O_2}$, consistent with \texttt{VULCAN}. The corresponding parameters are taken from multiple sources (Appendix A of \citealt{Tsai_2021}; \citealt{Marrero_1972, Moses2000_icarus, Banks1973, Hu_2012}).

Eddy diffusion is a complex physical process that is parameterized by the eddy diffusion coefficient $K_{\mathrm{zz}}$. Existing models in the community typically either (i) assume a constant $K_{\mathrm{zz}}$ value, chosen based on the type of planet, or (ii) obtain its vertical profile empirically or from a General Circulation Model (GCM).

We numerically discretize equations~\eqref{eq:continuity} and \eqref{eq:transportflux} following \citet{Hu_2012} and \citet{Rimmer_2016} as
\begin{equation} \label{eq: discrete_continuity}
	\begin{aligned}
		\frac{\partial n_{y,k}}{\partial t} &= P_{y,k} - L_{y,k} n_{y,k} 
		- \frac{d_{k+1/2} \left(\frac{n_{\mathrm{gas},k+1/2}}{n_{\mathrm{gas},k+1}}\right) n_{y,k+1}}{2} \\
		&\quad - \left[ d_{k+1/2} \left(\frac{n_{\mathrm{gas},k+1/2}}{n_{\mathrm{gas},k}}\right) - 
		d_{k-1/2} \left(\frac{n_{\mathrm{gas},k-1/2}}{n_{\mathrm{gas},j}}\right) \right] n_{y,k} \\
		&\quad + d_{k-1/2} \left(\frac{n_{\mathrm{gas},k-1/2}}{n_{\mathrm{gas},k}}\right) n_{y,k-1},
	\end{aligned}
\end{equation}
\vspace{0.3cm}

where $k$ denotes the current atmospheric layer, $k-1$ and $k+1$ represent the layers immediately below and above it, and $k \pm 1/2$ refers to the arithmetic mean of $k$ and $k \pm 1$. The diffusion coefficient term is given by
\begin{equation}\label{eq:dzz}
	d_{k\pm \frac{1}{2}} = \frac{D_{k\pm \frac{1}{2}}}{2\Delta z^2} \left[ \frac{(\overline{m} - m_y) g \Delta z}{k_B T_{k\pm \frac{1}{2}}} - \frac{\alpha_T}{T_{k\pm \frac{1}{2}}} \left(T_{k\pm 1} - T_k\right) \right],
\end{equation}
where $\overline{m}$ is the mean molecular mass of the atmosphere, and $m_y$, $P_{y,k}$, $L_{y,k}$, and $n_{y,k}$ are the molecular mass, production rate, loss rate, and number density, respectively, of the $y$th species at layer $k$.

Since eddy diffusion represents the bulk mixing of gases, the movement of the air parcel itself accounts for this process, and the eddy diffusion terms do not explicitly appear in the discretized equation. Following \citet{Rimmer_2016}, molecular diffusion is represented as a reaction in which the diffusing molecule is removed from the parcel and reintroduced via a reverse reaction, ensuring mass conservation.

Because the discretized continuity equation (equation~\ref{eq: discrete_continuity}) is a stiff ordinary differential equation (ODE), implicit solvers are the most suitable choice \citep{Tsai_2017, book}. Commonly used methods in the planetary chemical kinetics community include the lower-order backward Euler method \citep{Hu_2012}, the higher-order DLSODES solver \citep{Venot_2012, Rimmer_2016}, the Rosenbrock method \citep{Tsai_2017, Hobbs_2019, Tsai_2021}, and the CVODE BDF solver \citep{Wogan_2022, Wogan_2023}. Our model employs the \texttt{pylsodes} package\footnote{\url{https://github.com/kmaitreys/pylsodes}}, an in-house Python wrapper for the \texttt{DLSODES} solver from the \texttt{FORTRAN ODEPACK} family \citep{hindmarsh1982rs}. This package has been extensively used and benchmarked for astrochemical modeling involving large chemical networks \citep{pegasis_pylsodes}.

\subsection{The \texttt{XODIAC}--2025 Chemical Network and Newly Compiled \texttt{XODIAC} Thermochemical Database} \label{sec:xodnet}

\texttt{XODIAC} includes three chemical networks: \texttt{STAND-2020} \citep{Rimmer_2021}, the \texttt{SNCHO} network \citep{tsai2021comparative}, and \texttt{XODIAC-2025}, a newly developed, state-of-the-art chemical network built within the \texttt{XODIAC} framework. This flexibility enables comparative estimates of planetary atmospheric compositions within a single modeling environment using multiple chemical networks commonly adopted by the community.

The \texttt{XODIAC-2025} network was developed by incorporating:  
(i) 5720 unique reactions from the full \texttt{STAND-2020} network \citep{Rimmer_2021}, which treats \texttt{C-H-O-N-S-Metals} chemistry within a temperature range of 100--30,000~K,  
(ii) 1567 unique reactions involving simple and complex hydrocarbons from the \texttt{KINETICS} network proposed by \citet{willacy2022vertical},  
(iii) 390 unique reactions for phosphorus chemistry linking the PHO and PN networks \citep{lee2024photochemical,douglas2022experimental,Silva_2025},  
(iv) 24 updated reactions for sulfur chemistry \citep{tsai2024biogenic}, and  
(v) 19 updated reactions for sodium-bearing species \citep{acharyya2024formation}.  

The validity of the newly added 2000 reactions was tested within the same temperature range of 100--30,000~K by following the criterion proposed by \citet{Rimmer_2016}, which excludes rate constants that exceed the collisional limit at high temperatures. In total, the \texttt{XODIAC-2025} network comprises 7720 reactions involving 594 species, covering \texttt{C-H-O-N-S-Metals} chemistry over 100--30,000~K. Furthermore, the rate coefficients for individual reactions in \texttt{XODIAC-2025} were cross-checked against databases such as KIDA\footnote{\url{https://kida.astrochem-tools.org/}} and NIST\footnote{\url{https://kinetics.nist.gov/kinetics/}}, as well as with published chemical networks including \texttt{STAND-2020} \citep{Rimmer_2021} and \texttt{SNCHO} \citep{tsai2021comparative}, along with their original references.

% The evaluation followed the criteria proposed by \citet{Rimmer_2016} for updating and constructing a new reaction network for planetary atmospheres, after plotting the rate constants over a temperature range of 100--30,000~K:

% \begin{enumerate}
	%     \item For a given reaction, if only one published rate is available, that value is adopted.
	%     \item Rate constants that become unrealistically large at high temperatures, exceeding the collisional limit, are rejected.
	%     \item If multiple rate constants are available for the same reaction, the most recent value is selected. \textcolor{red}{If there are reactions common in the STAND2020 network and our newly added reactions, we consider the STAND network reactions only.}
	%     \item \textcolor{red}{If Arrhenius parameters of both forward and reverse reactions of any added reaction are available, we consider the reaction with the lower energy barrier as our forward reaction and use the NASA polynomials to reverse them thermodynamically.}
	% \end{enumerate}

Among the 594 species present in the \texttt{XODIAC-2025} network, only P($^2$D), \ce{P2O2}, and \ce{P2O} lack thermochemical data or NASA 7-coefficients in publicly available databases such as \texttt{NIST-JANAF}\footnote{\url{https://janaf.nist.gov/}}, \texttt{BURCAT}\footnote{\url{http://garfield.chem.elte.hu/Burcat/burcat.html}}, and \texttt{RMG}\footnote{\url{https://rmg.mit.edu/}}. For these three molecules, we computed the NASA 7-coefficients using \texttt{Gaussian~16} \citep{g16} in combination with the \texttt{Arkane} module of the \texttt{Reaction Mechanism Generator (RMG)} \citep{johnson2022rmg,liu2021rmg,dana2023_arkane} software (version~3.2.0), as detailed in Appendix~\hyperref[app:apdx]{A}. The remaining newly added species in the \texttt{XODIAC-2025} network that are not present in the \texttt{STAND-2020} network were primarily sourced from either the \texttt{BURCAT} or \texttt{RMG} databases, with the exception of N($^2$D), which was taken from \cite{Venot2012}. These species are summarized in Table~\ref{tab:nasa_sum} in Appendix~\hyperref[app:apdx]{A}.
%~\ref{app:apdx}.

\subsection{Types of reactions included in \texttt{XODIAC-2025} network and calculation of their rate coefficients} \label{subsec:rate-coefficient}

The \texttt{XODIAC-2025} network contains three broad categories of reactions: two-body reactions, three-body reactions, and photochemical reactions. The calculation of their rate coefficients is described in the following subsections.

\subsubsection{Two-Body Reactions}

Also called bimolecular reactions, these involve two reactants and are the most common reactions in atmospheric gas-phase chemistry. The rate coefficient for this type of reaction is second order, as it depends on collisions between two particles. Both neutral–neutral \eqref{eq:neutral-neutral} and ion–neutral \eqref{eq:ion-neutral} reactions fall under this category. A general framework for such reactions is given by

\begin{equation} \label{eq:neutral-neutral}
	C + D \rightarrow E + F,
\end{equation}
\begin{equation} \label{eq:ion-neutral}
	C^+ + D \rightarrow E^+ + F.
\end{equation}

The rate coefficients for these reactions are calculated using either the Kooij equation \citep{Kooij_1893} or the modified Arrhenius equation:

\begin{equation}
	k = \alpha \, T_b^{\beta} \, e^{(-\gamma/T_c)},
\end{equation}
where $T_b$ and $T_c$ are defined as follows:
\[
T_b =
\begin{cases} 
	\dfrac{T}{300}, & \text{(Kooij form)}, \\
	T, & \text{(modified Arrhenius form)},
\end{cases}
\quad \text{and} \quad T_c = T.
\]

Here, $k$ is the rate coefficient, $T$ is the gas temperature, and $\alpha$, $\beta$, and $\gamma$ are the Kooij/Arrhenius constants. These constants are obtained either from experimental databases or from theoretical calculations.

\subsubsection{Three-Body Reactions}

Also referred to as termolecular reactions, these processes are pressure dependent because their rates are governed by the efficiency of thermalization through collisions with a third body. Such reactions occur in both decomposition \eqref{eq:decomposition} and recombination \eqref{eq:recombination} processes, expressed as:

\begin{equation} \label{eq:decomposition}
	A + M \rightarrow B + C + M,
\end{equation}
\begin{equation} \label{eq:recombination}
	B + C + M \rightarrow A + M,
\end{equation}
where $M$ denotes a third particle that remains chemically unaffected. In decomposition reactions, $M$ acts as an energy source, while in recombination reactions, it serves as an energy sink.  

For these reactions, the rate coefficient is calculated using the Lindemann form \citep{Lindemann1922}, with expressions for both the low- and high-pressure limits:
\begin{equation}
	k_0 = \alpha_0 \, T_b^{\beta_0} \, e^{-(\gamma_0/T)},
\end{equation}
\begin{equation}
	k_{\infty} = \alpha_{\infty} \, T_b^{\beta_{\infty}} \, e^{-(\gamma_{\infty}/T)},
\end{equation}
where $k_0$ and $k_{\infty}$ are the low- and high-pressure rate coefficients, respectively.  

Treating the process as an effective bimolecular reaction, the overall rate coefficient for the three-body reaction is given by:
\begin{equation}
	k = \frac{k_0 [M]}{1 + \frac{k_0}{k_{\infty}} [M]},
\end{equation}
where $[M]$ is the number density of the third body (which may be any species present in the atmosphere).

\subsubsection{Photochemical Reactions} \label{subsubsec:photoreact}
Photochemistry represents a fundamental class of reaction mechanisms in any atmospheric chemical model and therefore requires flexibility in treating photochemical processes by considering the available photo–cross-section data. As explained in detail below, \texttt{XODIAC} is designed to handle both reaction-specific and molecule-specific photo–cross-section data, dividing them into uniform or non-uniform wavelength bins.

When solar radiation passes through a planetary atmosphere, photons interact with atmospheric molecules, particularly in the upper layers where ultraviolet (UV) radiation is most intense. These molecules absorb the incoming UV photons, and depending on the photon energy, may undergo photo-dissociation to form reactive radicals. The rate of photochemical reactions is directly proportional to the number density of the reacting species at a given atmospheric level.
The emergent stellar radiation carries energy across a spectrum of wavelengths capable of driving photochemical processes in planetary atmospheres. This energy is typically represented in terms of the spectral actinic flux, $\mathrm{J(z, \lambda)}$, defined as the number of photons incident from all directions per unit time, per unit area, and per unit wavelength, where \( z \) denotes the atmospheric altitude and $\mathrm{\lambda}$ is the wavelength.

In \texttt{XODIAC}, $\mathrm{J(z, \lambda)}$ can be computed using two distinct components. The first is the diffusive flux,  $\mathrm{J_{\text{diff}}}$, which accounts for the contribution of multiply scattered radiation within the atmosphere. It is expressed as:

\[
J_{\text{diff}} = \frac{1}{2} \left( J\uparrow + J\downarrow \right)
\]

where $\mathrm{J\uparrow}$  and $\mathrm{J\downarrow}$  represent the upward and downward diffusive fluxes, respectively. The second component corresponds to the direct stellar radiation attenuated by atmospheric extinction. Therefore, the total actinic flux at a given altitude is given by:

\[
J(z, \lambda) = J_0 \, e^{-\tau(z, \lambda)/\mu} + J_{\text{diff}} 
\]

where $\mathrm{J_0}$ is the incoming stellar flux at the top of the atmosphere,  $\mathrm{\tau(z, \lambda)}$ is the optical depth, and $\mathrm{\mu}$ is the cosine of the zenith angle of the incoming radiation. The observed spectral flux has been taken from the sources like \texttt{PHOENIX} \footnote{\url{https://archive.stsci.edu/hlsps/reference-atlases/9cdbs/grid/phoenix/}} and \texttt{ATLAS} \footnote{\url{https://archive.stsci.edu/hlsps/reference-atlases/cdbs/grid/ck04models/}} databases. For using the stellar spectra, one needs to re-normalize the incoming stellar to the TOA of the planet by scaling with $(r/a)^2$, where $r$ is the planetary radius and $a$ is the star-planet separation. \texttt{XODIAC} is currently equipped to handle both uniform and non-uniform spectral flux distributions, allowing spectral flux to be defined over arbitrarily spaced wavelength intervals as required by the specific scientific context. 

For calculating direct radiation at each pressure level, we need the corresponding optical depths ($\mathrm{\tau}$), which includes the contribution of molecular absorption ($\mathrm{\tau_{a}}$) and Rayleigh scattering ($\mathrm{\tau_{r}}$). This optical depth $\mathrm{\tau}$, which represents the total extinction, is given by:

\[
\tau = \tau_{\text{a}} + \tau_\text{r} = \left[ \sum_i (\sigma_{a,i} + \sigma_{s,i}) \ n_i \ dz \right] \,
\]

Here, $\mathrm{\sigma_{a,i}}$ and $\mathrm{\sigma_{s,i}}$ represent the absorption and scattering cross sections, respectively. It is important to note that $\mathrm{\sigma_{a,i}}$ may differ from the photo-dissociation cross section, as photon absorption does not always result in molecular dissociation. Accordingly, in the current version of \texttt{XODIAC}, the optical depths can be computed using either photo-absorption or photo-dissociation cross-section data, depending on the modeling context.

\texttt{XODIAC} primarily utilizes cross-section data from the \texttt{PHIDRATES} database~\citep{huebner1992ap,huebner2015photoionization}\footnote{\url{https://phidrates.space.swri.edu}} when available. It is also compatible with cross-section data sourced directly from the Leiden Observatory database~\citep{heays2017photodissociation}\footnote{\url{https://home.strw.leidenuniv.nl/~ewine/photo/cross_sections.html}}, which has been well benchmarked against other widely recognized databases such as \texttt{PHIDRATES}. It provides tabulated photo-absorption, photo-dissociation, and photo-ionization cross-sections, along with associated uncertainty classifications. All the relevant Rayleigh scattering data have been taken from \cite{sneep2005direct} and \cite{thalman2014rayleigh}. 

In the current version of \texttt{XODIAC}, the photo cross-sections have been divided into 10{,}000 wavelength bins, each approximately $1\,\mathrm{\AA}$ wide. The wavelength range used for scientific analysis spans from \(\lambda_{\text{min}} = 1\,\text{\AA}\) to \(\lambda_{\text{max}} = 10{,}000\,\text{\AA}\). However, \texttt{XODIAC} is designed with flexibility, allowing the specification of arbitrary wavelength ranges and bin resolutions tailored to the requirements of a given analysis (e.g., see section \ref{subsec: Benchmarking with VULCAN} where the entire wavelength range is non-uniformly discretized). 
% further demonstrating the adaptability of the code.

The \texttt{PHIDRATES} database provides reaction-specific photo-cross section data, which can be directly employed to compute the corresponding photochemical rate coefficients. In contrast, the Leiden Observatory database offers molecule-specific cross section datasets. In such cases, obtaining the effective photo-cross section for a particular photochemical reaction requires scaling the molecular cross section by the quantum yield $\mathrm{q(\lambda)}$ [photons\(^{-1}\)] after interpolating it for the specific wavelength range, which represents the probability of a specific photo-reaction occurring upon photon absorption. \texttt{XODIAC} can handle both types of cross-section data, reaction-specific and molecule-specific, ensuring compatibility with diverse datasets and maintaining flexibility in photochemical modeling.

% \[
% \sigma_i(\lambda) = \sigma_{PHIDRATES,\,i}(\lambda)\ \text{or},\ \sigma_{i}(\lambda)=\sigma_{Leiden}(\lambda) \,q_{i}(\lambda)
% \]

\[
\sigma_i(\lambda) = 
\left\{
\begin{array}{l}
	\sigma_{\text{PHIDRATES},\,i}(\lambda) \quad \text{(for PHIDRATES)} \\
	\sigma_{\text{Leiden}}(\lambda) \, q_i(\lambda) \quad \text{(for Leiden)}
\end{array}
\right.
\]

where, $i$ corresponds to the specific photo-chemical reaction.  

Finally, tabulated chemical cross sections are integrated with the actinic flux $\mathrm{J(z, \lambda)}$ [cm\(^{-2}\) s\(^{-1}\) \AA\(^{-1}\)], which represents the radiant flux density incident on a unit sphere located at an atmospheric height z [cm], to compute the photochemical rate constants. For a photochemical reaction, say

\[
\mathrm{A} \xrightarrow{h\nu} \mathrm{B} + \mathrm{C}
\]
these rate constants are evaluated using
\[
k_{\text{ph}, i}(z) = t_f \,\int_{\lambda_{min} }^{\lambda_{max}} J(z, \lambda)\, \sigma_{i}(\lambda)\, d\lambda
\]

and the photolysis rate of that photochemical reaction is given by:

\[
\frac{dn_A}{dt} = -k_{\text{ph}, i}\ n_A
\]

where the index $i$ runs over the molecular species considered in the photochemical network. The parameter $\mathrm{t_f}$ is a dimensionless scaling factor that accounts for the fraction of time a particular atmospheric region is irradiated over timescales much longer than the atmospheric dynamical timescale. For tidally locked planets, $\mathrm{t_f}$ = 1 on the day side and $\mathrm{t_f}$ = 0 on the night side. For planets with regular rotation, the diurnally averaged value is typically $\mathrm{t_f}$ = $\mathrm{\frac{1}{2}}$. 

%---------------------------------------------------------

\subsubsection{Reverse Reactions}
The rate coefficients of all reverse reactions in the \texttt{XODIAC-2025} network are consistently calculated using the thermodynamical data compiled within the \texttt{XODIAC} thermochemical database.

For a forward reaction of the form
\begin{equation} \label{eq: forward}
	A + B \rightarrow X + Y + Z
\end{equation}
and its corresponding reverse reaction
\begin{equation} \label{eq: reverse}
	X + Y + Z \rightarrow A + B,
\end{equation}
the rate coefficients for the forward reaction, $k_f$ \eqref{eq: forward}, and the reverse reaction, $k_r$ \eqref{eq: reverse}, are related to the equilibrium constant $k_c$ by
\begin{equation}
	k_c = \frac{k_f}{k_r}.
\end{equation}

The equilibrium constant is obtained from the Gibbs free energy of the reaction (see Appendix~E of \cite{Tsai_2017} and \cite{burcat_2005}) and is then used directly to calculate the rate coefficients of the reverse reactions.

\begin{figure}
	\centering
	\includegraphics[width=\linewidth]{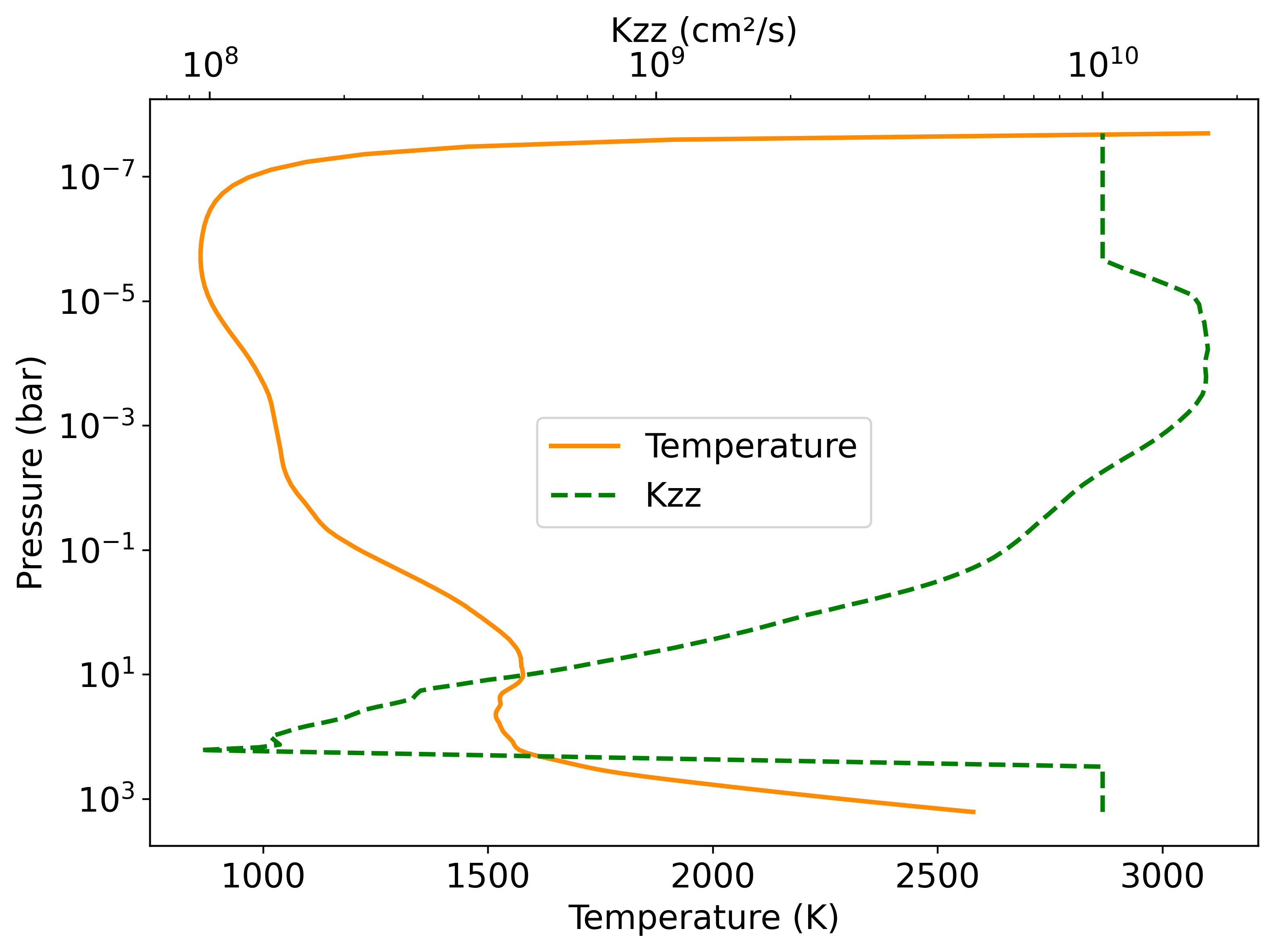}%{K_zz.png}%
	\caption{Pressure–temperature and eddy diffusion coefficient ($K_{\mathrm{zz}}$) profile of HD~189733~b (adopted from \citealt{Mosses_2011})}.
	
	\label{fig: P_T_profile}
\end{figure}

\subsection{Atmospheric Escape}\label{subsec:escape}

\texttt{XODIAC} also considers atmospheric escape processes, following the approach of \cite{Hu_2012}, in which three scenarios are implemented: (i) no escape, (ii) escape of only H and He, and (iii) escape of any molecule based on user-defined conditions. The third case, where any species can escape, is very rare; typically, either atmospheric escape is neglected or it is considered only for hydrogen and helium. For the second case, the diffusion-limited flux (\cite{Hu_2012}; \cite{Tsai_2021}) is given by
\begin{equation}
	V_{\text{lim}} = D_{k\pm\frac{1}{2}} \left( \frac{1}{H_0} - \frac{1}{H} \right),
\end{equation}
where $D_{k\pm\frac{1}{2}}$ is the molecular diffusion coefficient from equation~\eqref{eq:dzz}, and $V_{\text{lim}}$ is the maximum escape flux in the upper atmosphere.

\subsection{Initial and Boundary Conditions} \label{subsec: ini condition}

Initial conditions play a major role in determining both the starting point and the steady state of an atmosphere \citep{Tsai_2021}. In \texttt{XODIAC}, the initial atmospheric composition can be specified using one of three options:  
(i) equilibrium abundances from the built-in equilibrium chemistry solver \texttt{NEXOCHEM} \citep{deka2025nexotrans},  
(ii) equilibrium abundances from the open-source code \texttt{FASTCHEM} \citep{Stock2022}, or  
(iii) user-defined abundances.

Boundary conditions also play a crucial role in studying the compositional changes that can occur in an exoplanetary atmosphere over the timescale of atmospheric evolution. For terrestrial exoplanets, the lower boundary condition may constrain surface--atmosphere interactions and interior composition (through outgassing/surface emission or deposition rates), while the upper boundary condition may determine atmospheric mass loss or gain (through escape or capture processes) \citep{Hu_2012}. However, for hot Jupiters, the presence of a surface is uncertain, and the lower boundary condition is typically set by chemical equilibrium \citep{Tsai_2021}. Our model currently includes the option to specify upper boundary conditions (\ref{subsec:escape}); however, as we are presently focused on benchmarking the model with hot Jupiters, we restrict any exchange of material across the boundaries \citep{Tsai_2021,Mosses_2011,Venot_2012,Rimmer_2016}. Therefore, we adopt zero-flux boundary conditions at both the top and bottom of the atmosphere.

\begin{figure*}[htbp]
	\centering
	% top row
	\includegraphics[width=0.48\textwidth]{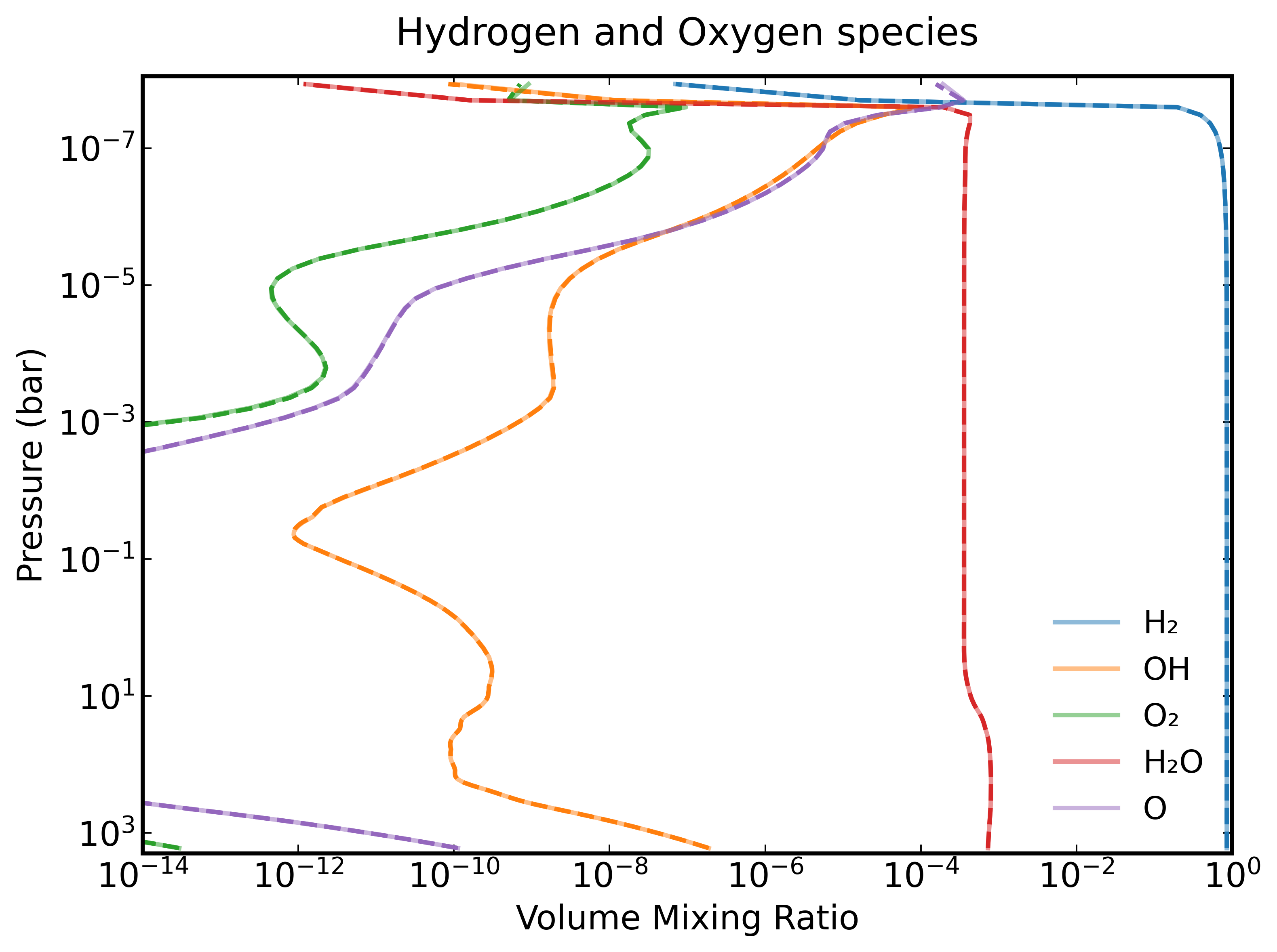}
	\hfill
	\includegraphics[width=0.48\textwidth]{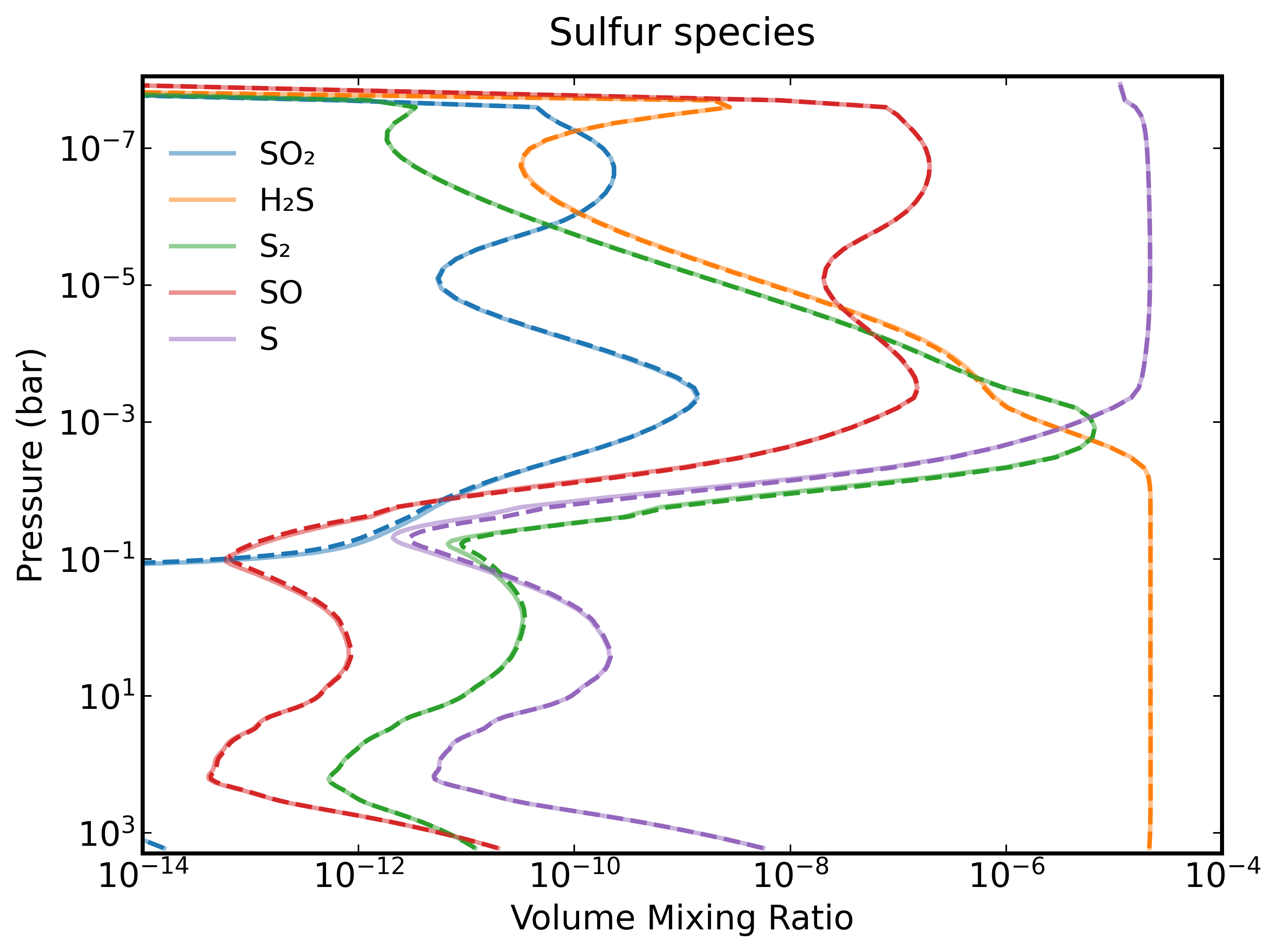}
	
	\vskip\baselineskip
	
	% bottom row
	\includegraphics[width=0.48\textwidth]{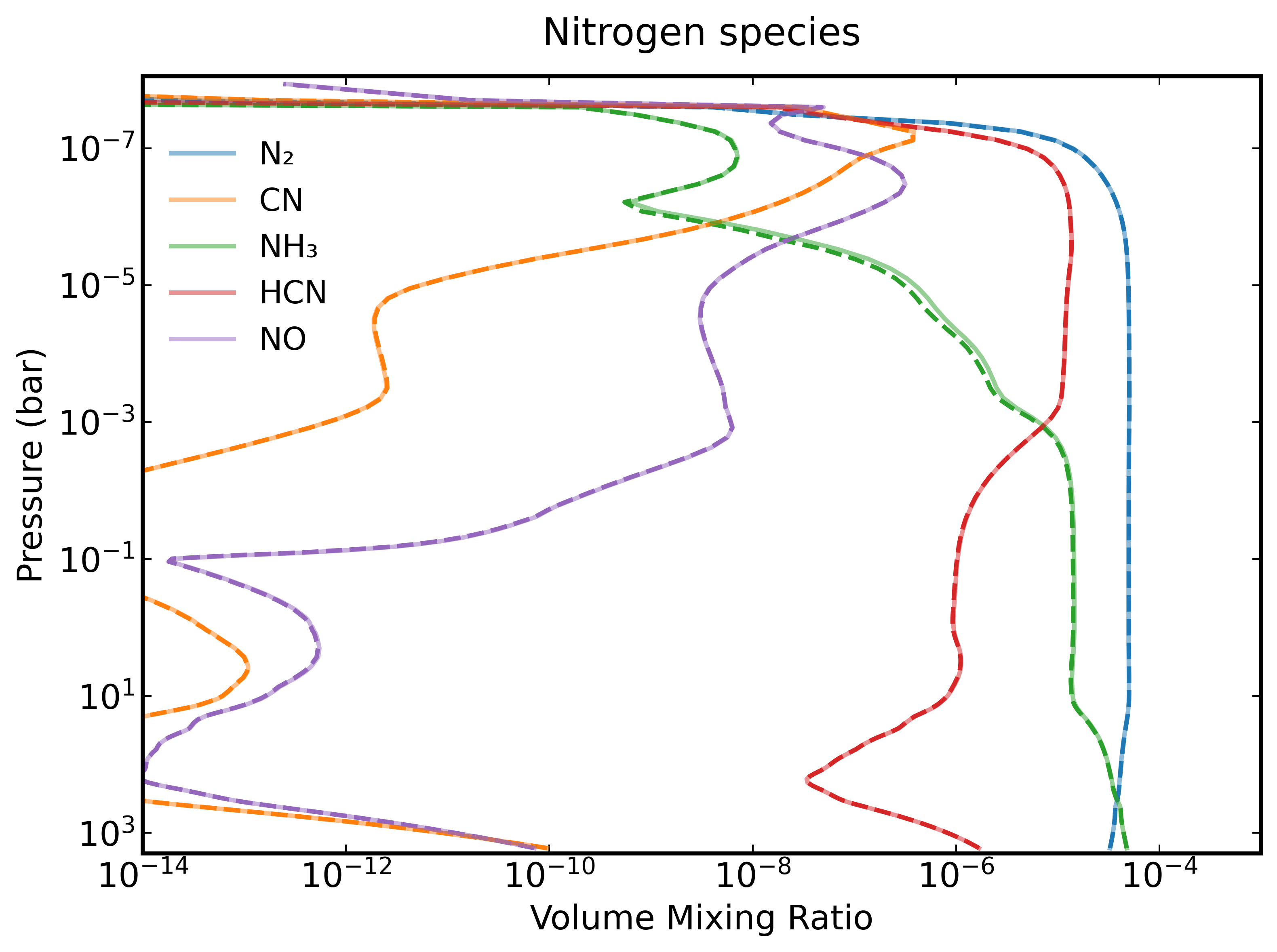}
	\hfill
	\includegraphics[width=0.48\textwidth]{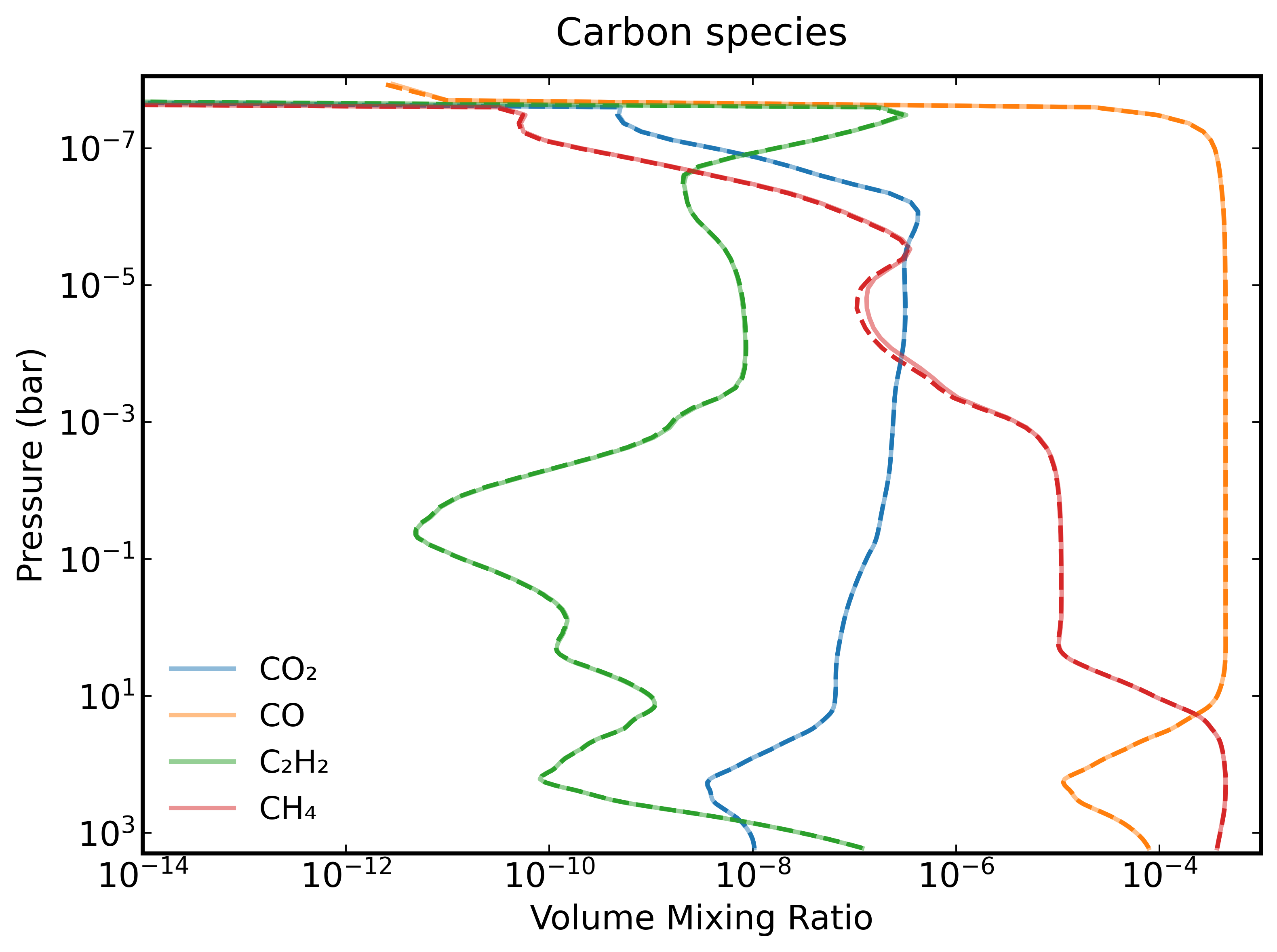}
	
	\caption{Comparison between \texttt{XODIAC} (solid lines) and \texttt{ARGO} (dotted lines)
		using the STAND-2020 reaction network for HD~189733b, showing species containing S, N, C, H, and O.}
	\label{fig:argo_benchmark}
\end{figure*}

\section{Validation and Performance Test of \texttt{XODIAC}} \label{subsec: XODIAC Validation}
To validate our model against established photochemical codes widely used in the community, we benchmarked it against \texttt{ARGO} \citep{Rimmer_2016} and \texttt{VULCAN} \citep{Tsai_2021} for HD~189733b, adopting the pressure–temperature and eddy diffusion coefficient ($K_{zz}$) profiles from \cite{Mosses_2011}, with initial abundances set to solar values. This benchmarking enables us to assess \texttt{XODIAC}’s capability to simulate atmospheric chemistry by directly comparing its outputs with those from models that employ distinct numerical approaches: the Lagrangian formalism, in which the atmosphere is evolved by moving a 1-D parcel through all pressure layers (\texttt{ARGO}); and the Eulerian formalism, in which the atmosphere is divided into a fixed pressure grid and all grid points evolve simultaneously (\texttt{VULCAN}).

\begin{figure*}[htbp]
	\centering
	% top row
	\includegraphics[width=0.48\textwidth]{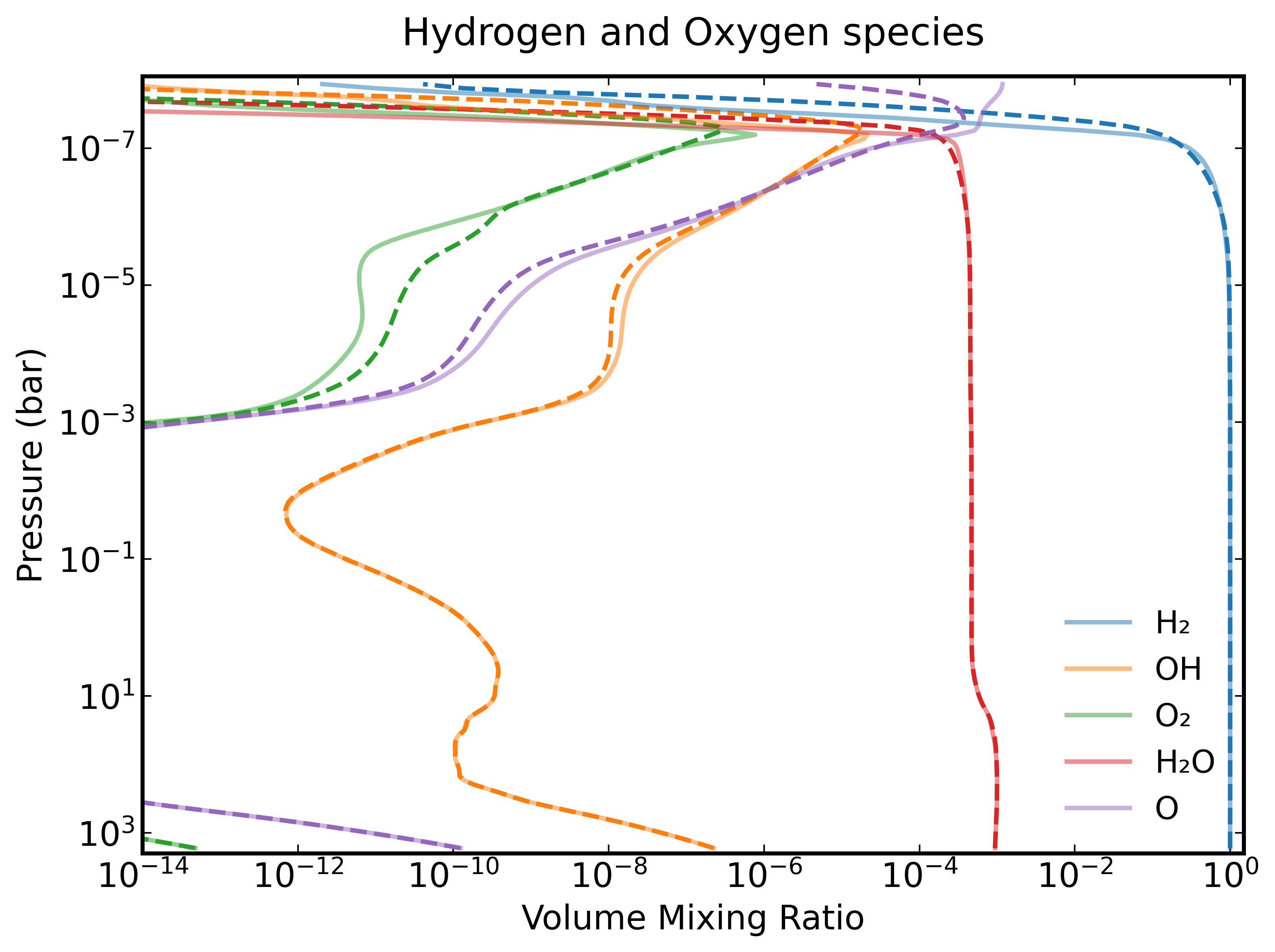}
	\hfill
	\includegraphics[width=0.48\textwidth]{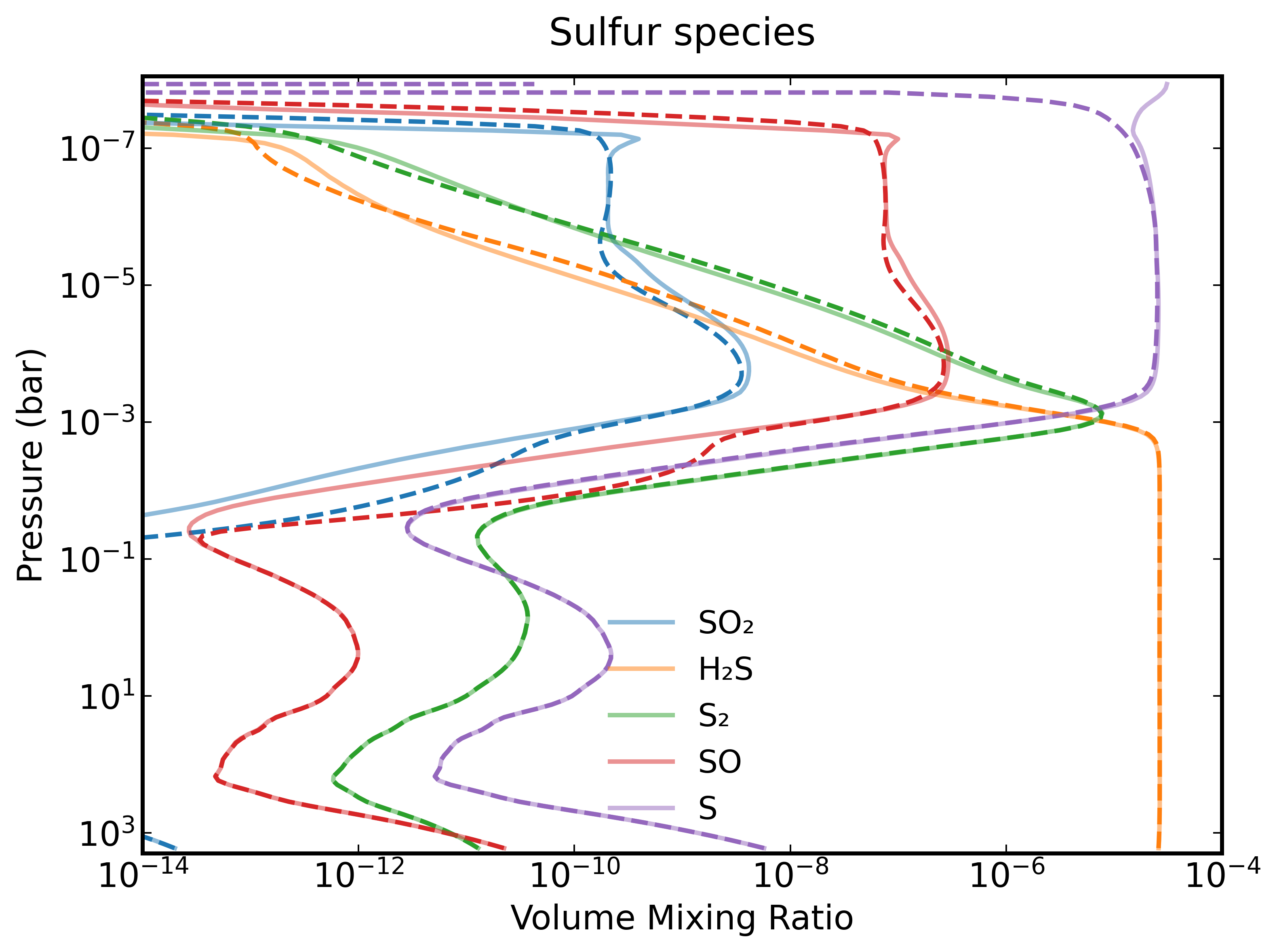}
	
	\vskip\baselineskip
	
	% bottom row
	\includegraphics[width=0.48\textwidth]{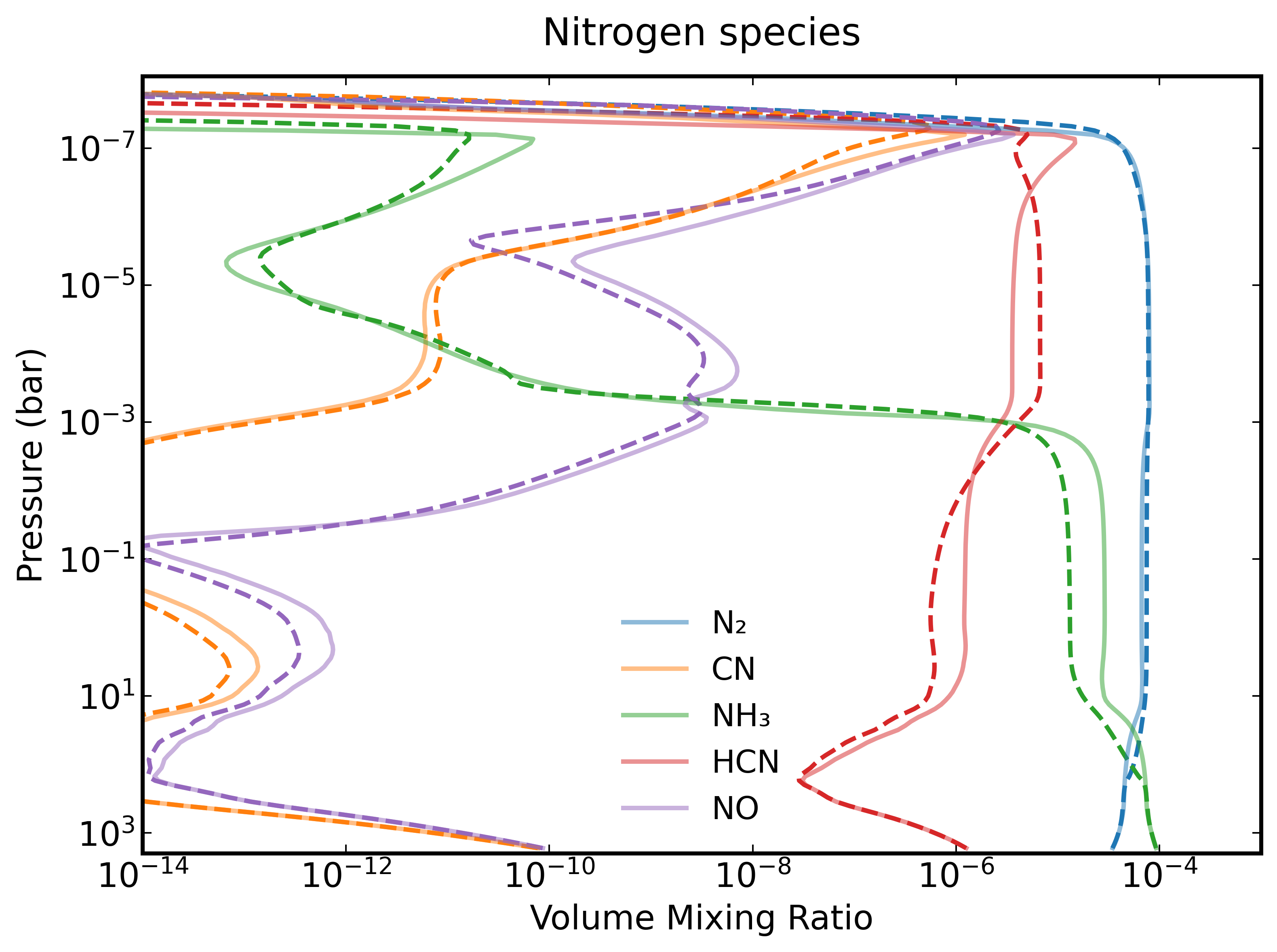}
	\hfill
	\includegraphics[width=0.48\textwidth]{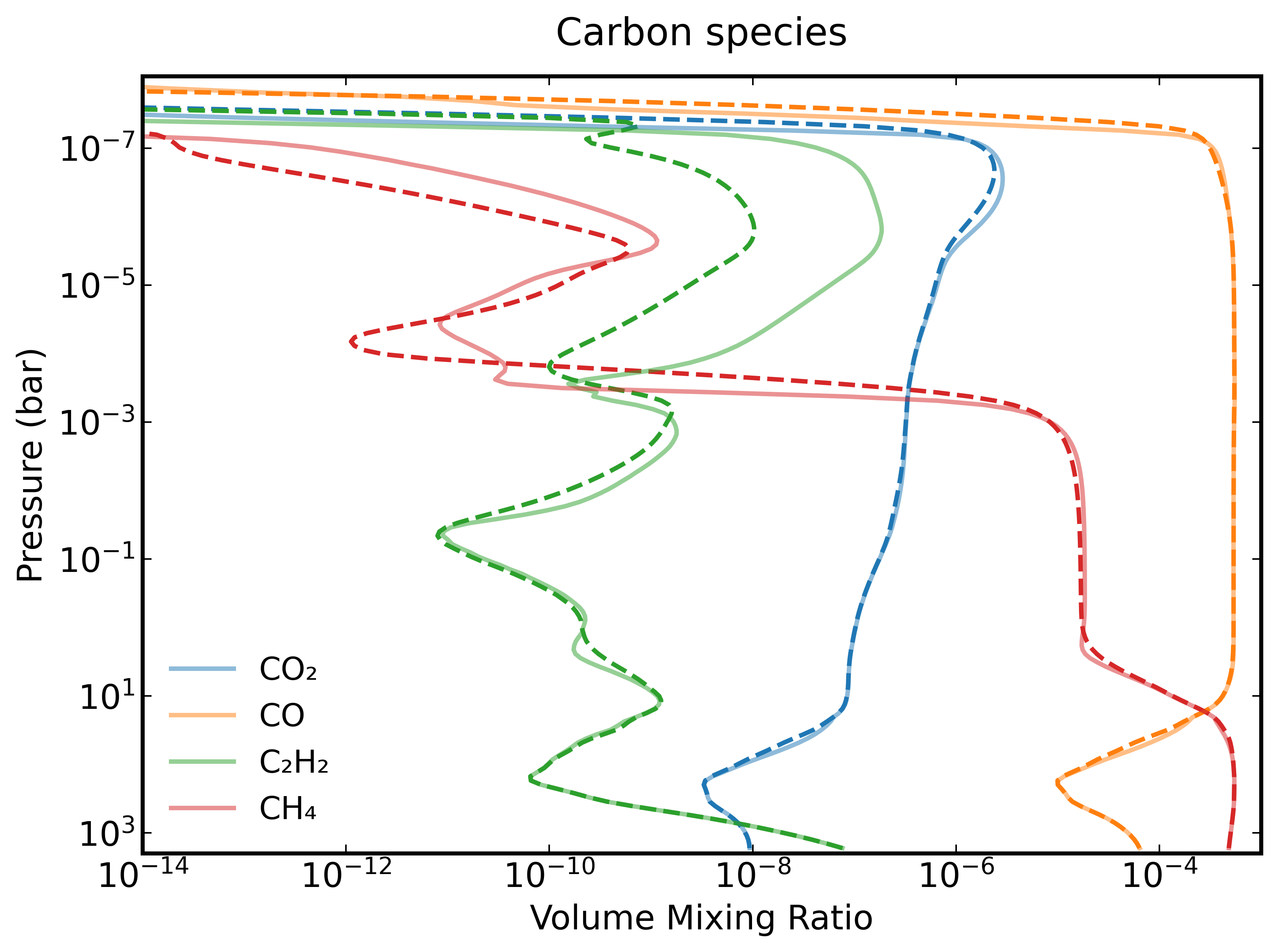}
	
	\caption{Comparison between \texttt{XODIAC} (solid lines) and \texttt{VULCAN}
		(dotted line) with the VULCAN-SNCHO reaction network for HD~189733b, showing
		species containing S, N, C, H, and O.}
	\label{fig:vulcan_benchmark}
\end{figure*}

We perform three benchmarking exercises:
\begin{enumerate}
	\item[(i)] Comparing \texttt{XODIAC} results using the \texttt{ARGO} \texttt{STAND-2020} network (\texttt{XODIAC-STAND2020}) with \texttt{ARGO} results using the same \texttt{STAND-2020} network (\texttt{ARGO-STAND2020}).
	\item[(ii)] Comparing \texttt{XODIAC} results using the \texttt{VULCAN} \texttt{SNCHO} network (\texttt{XODIAC-SNCHO}) with \texttt{VULCAN} results using the same \texttt{SNCHO} network (\texttt{VULCAN-SNCHO}).
	\item[(iii)] Evaluating the computational performance of \texttt{XODIAC} with three different networks, \texttt{XODIAC-2025}, \texttt{XODIAC-STAND2020}, and \texttt{XODIAC-SNCHO}, in comparison with \texttt{ARGO} (\texttt{ARGO-STAND2020}) and \texttt{VULCAN} (\texttt{VULCAN-SNCHO}).
	
\end{enumerate}

\subsection{Benchmarking of \texttt{XODIAC-STAND2020} with \texttt{ARGO-STAND2020} network} \label{subsec: Benchmarking with ARGO}
	
	To validate our implementation, we performed rigorous benchmarking against \texttt{ARGO} \citep{Rimmer_2016}, a well-established atmospheric chemistry model employing a Lagrangian framework. We reconstructed the \texttt{ARGO} chemical network in the \texttt{XODIAC} format, referring to it as the \texttt{XODIAC-STAND2020} network. This network contains 5,720 reactions, encompassing two-body, three-body, and photochemical processes. We adopted the same thermochemical database used in \texttt{ARGO} and implemented the photochemical cross-sections for 209 photochemical branches, sourced from the \texttt{PHIDRATES} database. Stellar fluxes were interpolated over $1$–$10^4~\mathrm{\AA}$ using surface fluxes from the \texttt{ATLAS} stellar spectral library, scaled to the top of the atmosphere (TOA) by the factor $(r/a)^2$, where $r$ is the stellar radius and $a$ is the orbital distance. All relevant stellar and planetary parameters were obtained from the Exo.MAST\footnote{\url{https://exo.mast.stsci.edu/exomast_planet.html?planet=hd189733b}} database. The benchmarking results are shown in Figure~\ref{fig:argo_benchmark}. Our simulations show excellent agreement with \texttt{ARGO}, with quantitative comparisons confirming that key chemical features and trends are accurately reproduced. This validates both the robustness and accuracy of our implementation.
	
	\begin{figure*}[ht]
		\centering
		\includegraphics[width=\linewidth]{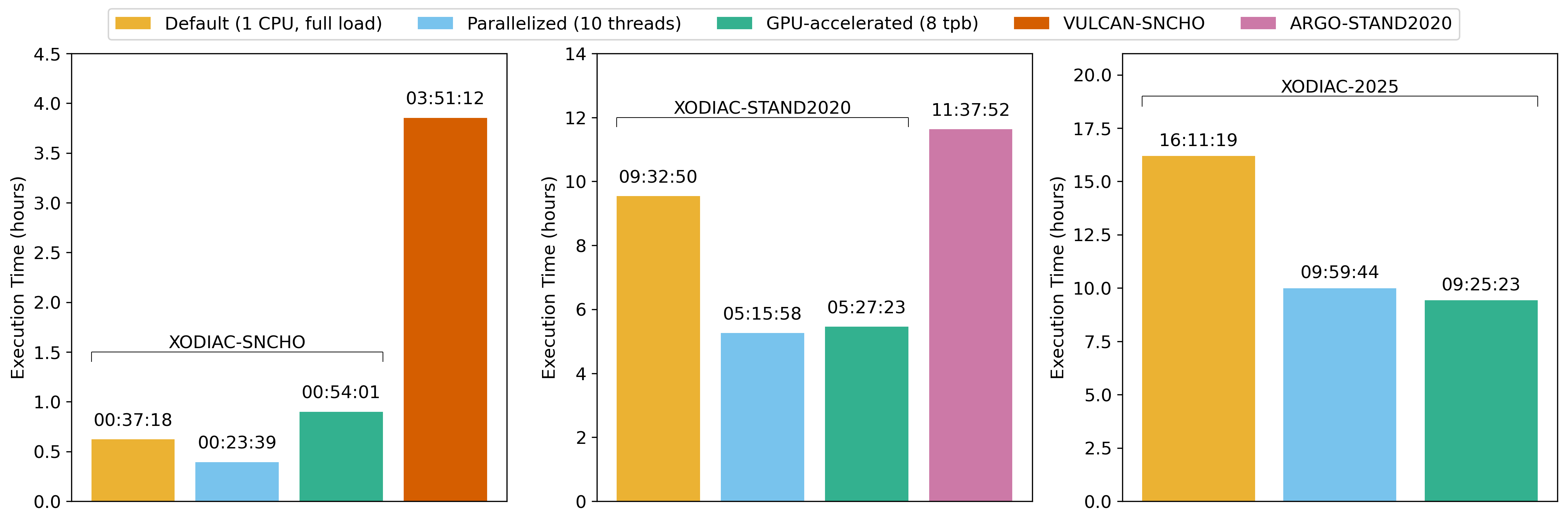}
		\caption{Performance comparison of three configurations of \texttt{XODIAC}: (1) default configuration (1 CPU, 100\% usage), (2) parallelized with Numba (10 threads), and (3) GPU-accelerated (8 threads per block). The exact runtime in \texttt{hh:mm:ss} format is displayed above each bar. For reference, runtimes from \texttt{VULCAN-SNCHO} and \texttt{ARGO-STAND2020} are shown alongside \texttt{XODIAC-SNCHO} and \texttt{XODIAC-STAND2020}.}
		\label{fig:speedtest}
	\end{figure*}
	
	\subsection{Benchmarking of \texttt{XODIAC-SNCHO} with \texttt{VULCAN-SNCHO} network}\label{subsec: Benchmarking with VULCAN}
	
	We benchmarked \texttt{XODIAC} against \texttt{VULCAN}, a widely used open-source photochemical kinetics model in the exoplanet community, to evaluate whether differences in computational schemes significantly influence simulation outcomes. To ensure consistency, we constructed an identical chemical network in the \texttt{XODIAC} format based on the \texttt{VULCAN-SNCHO} network \citep{Tsai_2021}, hereafter referred to as the \texttt{XODIAC-SNCHO} network. 
	
	For stellar spectra, we used the observed UV flux of HD~189733 from \citet{Bourrier2020}, who derived a semi-synthetic UV spectrum by combining observations from \textsc{HST} and \textsc{XMM-Newton}. This stellar spectrum was adopted from the \texttt{VULCAN} repository and scaled to the top of the atmosphere (TOA) of HD~189733~b. Photochemical cross-sections were obtained from the Leiden Observatory database\footnote{\url{https://home.strw.leidenuniv.nl/~ewine/photo/cross_sections.html}}, which provides photodissociation and photoabsorption cross-sections, along with branching ratios, for all photochemically active species included in \texttt{VULCAN}. The calculation of fluxes, including optical depth effects, is described in Section~\ref{subsubsec:photoreact}. 
	
	Following \texttt{VULCAN}, we used two wavelength ranges with different bin widths: $20$--$2400~\mathrm{\AA}$ with a bin width of $1~\mathrm{\AA}$, and $2400$--$7000~\mathrm{\AA}$ with a bin width of $20~\mathrm{\AA}$, for both the photochemical cross-sections and the stellar flux. Ensuring that the wavelength grids of the cross-sections and stellar flux are aligned is essential for accurate results. Figure~\ref{fig:vulcan_benchmark} compares mixing ratio profiles generated using the \texttt{VULCAN-SNCHO} and \texttt{XODIAC-SNCHO} networks. Our results show strong agreement with those from \texttt{VULCAN}, with most species matching closely. The primary exception is $\mathrm{C_2H_2}$, which shows slightly larger deviations at low-pressure regions.
	
	The temporal evolution in our model requires the variation timescale of each species at each layer to be larger than the mixing timescale of that layer until it reaches an evolution time set by the user (see \cite{Hu_2012}). In our model, variation timescales are applied uniformly across all species at each atmospheric layer. This can affect the predicted abundances of hydrocarbons, such as $\mathrm{C_2H_2}$, which are strongly influenced by their respective quench levels. The quenching approximation can yield solutions that deviate significantly beyond the uncertainties associated with dynamical timescales, as discussed by \cite{Tsai_2017}. Since individual species may quench at different depths, interactions with non-quenched species can alter the abundances of quenched species. Once these deviations occur, even strong incident stellar radiation may not be able to replenish their abundances via photochemical reactions.
	
	\subsection{Performance test of \texttt{XODIAC} compared to \texttt{ARGO} and \texttt{VULCAN}} \label{subsec:efficiency}
	The output of successive iterations in Lagrangian-based photochemical models depends on the results from preceding time steps and atmospheric layers. Consequently, such models typically require substantial computational time. In our implementation, the calculation of rate coefficients constitutes the most time-intensive component. The most effective strategy to accelerate these calculations is to perform them in parallel by distributing the workload across different memory units.
	
	To achieve this, we employed Numba’s just‑in‑time (JIT) compilation framework using \texttt{@njit} and \texttt{@cuda.jit} decorators to significantly accelerate execution by distributing rate coefficient computations across multiple CPU and GPU threads. We benchmarked the execution times for 10 global iterations using the T–P profile from Figure~\ref{fig: P_T_profile}, which consists of 182 layers, for three networks: \texttt{XODIAC-SNCHO}, \texttt{XODIAC-STAND2020}, and \texttt{XODIAC-2025} (see Section~\ref{subsec: XODIAC Validation}), on a Linux system equipped with an Intel\textsuperscript{\textregistered} Core\texttrademark~i7-10700 CPU @ 2.90~GHz and an NVIDIA GeForce GTX 1050 Ti GPU. These were compared with \texttt{ARGO} (using \texttt{ARGO-STAND2020}) and \texttt{VULCAN} (using \texttt{VULCAN-SNCHO}). GPU acceleration was implemented via a custom \texttt{CUDA} kernel written with Numba, configured with 8 threads per block (tpb). Although this configuration does not fully utilize the GPU's potential, with usage remaining below 10\%, it is expected to scale efficiently for networks significantly larger than \texttt{XODIAC-2025}. The current implementation is therefore scalable and well suited for future studies involving substantially larger networks.

	In our GPU-accelerated framework, both the GPU and CPU (for certain reactions) are employed to reduce runtime. For smaller networks, such as \texttt{XODIAC-SNCHO}, CPU-based parallelization with Numba (10 threads) outperforms the GPU approach due to the overhead from CPU–GPU memory transfers. However, for larger networks, the GPU achieves superior performance even at low \texttt{tpb} values. A comparative analysis of model execution times under various parallelization strategies is presented in Figure~\ref{fig:speedtest}. In all cases shown, \texttt{XODIAC} demonstrates higher computational efficiency than \texttt{ARGO} and \texttt{VULCAN}.
	
	To the best of our knowledge, this represents the first attempt within the exoplanet chemical kinetics community to fully exploit both CPU and GPU parallelization. This optimization yields substantial performance gains, reducing the total runtime by several hours for larger networks.
	
	\section{Comparative atmospheric chemistry of HD~189733~b across three chemical networks}\label{sec:res} 
	
	In this section, we revisit the atmospheric chemistry of the hot Jupiter HD~189733~b using \texttt{XODIAC} with three built-in chemical networks: \texttt{XODIAC-SNCHO}, \texttt{XODIAC-STAND2020}, and \texttt{XODIAC-2025}. This allows us to perform a comparative study of atmospheric chemistry across the three networks using the same photochemical model, while also providing a fresh perspective from the state-of-the-art \texttt{XODIAC-2025} network. Our analysis focuses on the formation and destruction pathways of key carbon-, sulfur-, and phosphorus-bearing molecules. 
	
	To better understand the underlying chemistry, we developed a reaction pathway analyzer designed to identify the most efficient conversion routes between any two species. This tool is inspired by the approach of \citet{Tsai_2018} and is built upon Dijkstra's classic algorithm \citep{Dijkstra1959}, which identifies the kinetically fastest reaction pathway by minimizing the cumulative sum of inverse reaction rates. Using this method, we systematically mapped the principal chemical pathways that govern the abundances of major species within the planet's atmosphere.

	\subsection{Comparative Carbon Chemistry}\label{subsec: carbon chemistry}
	
	\begin{figure*}[htbp]
		\centering
		
		\includegraphics[width=0.48\textwidth]{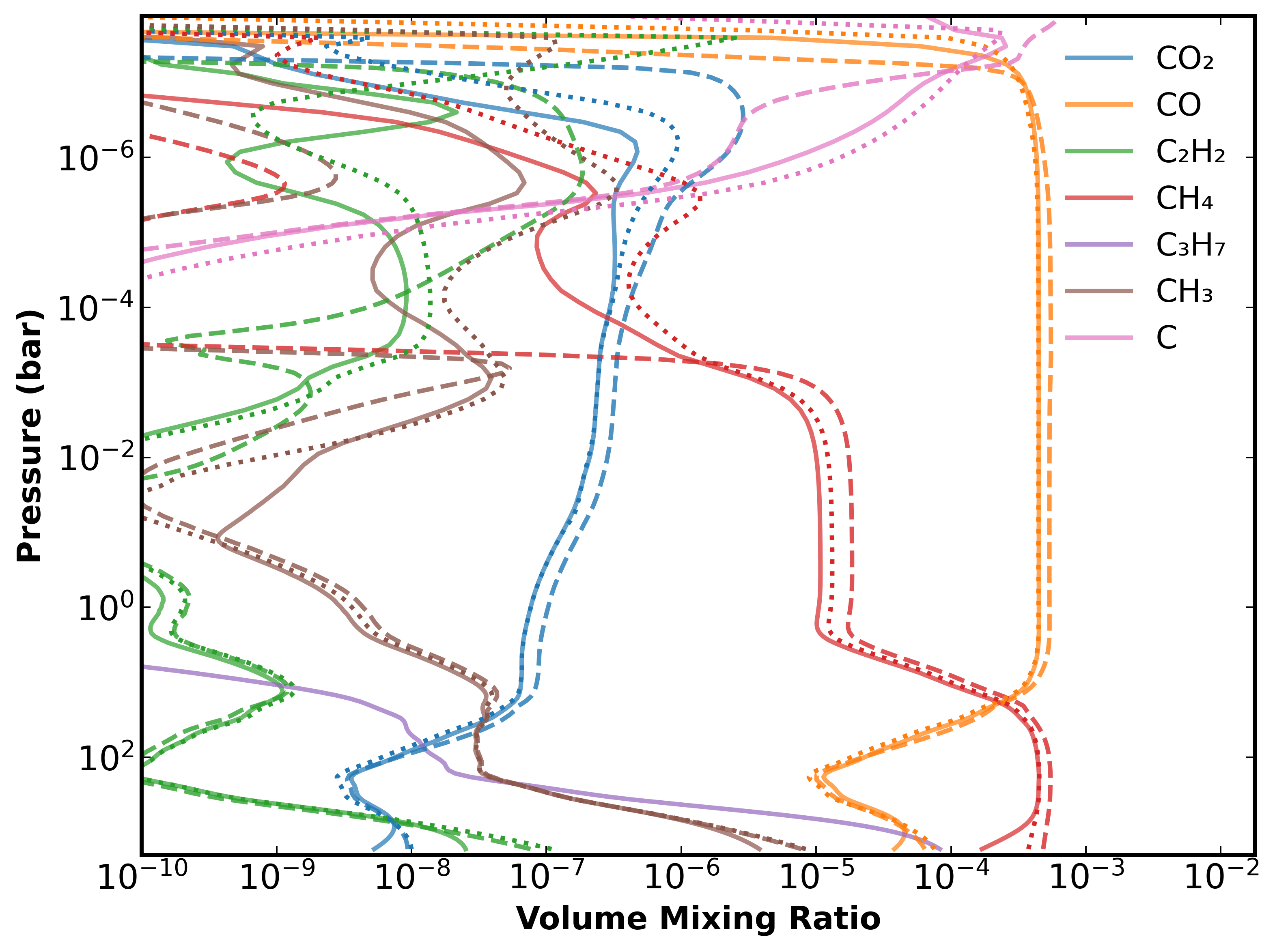}
		\hfill
		\includegraphics[width=0.48\textwidth]{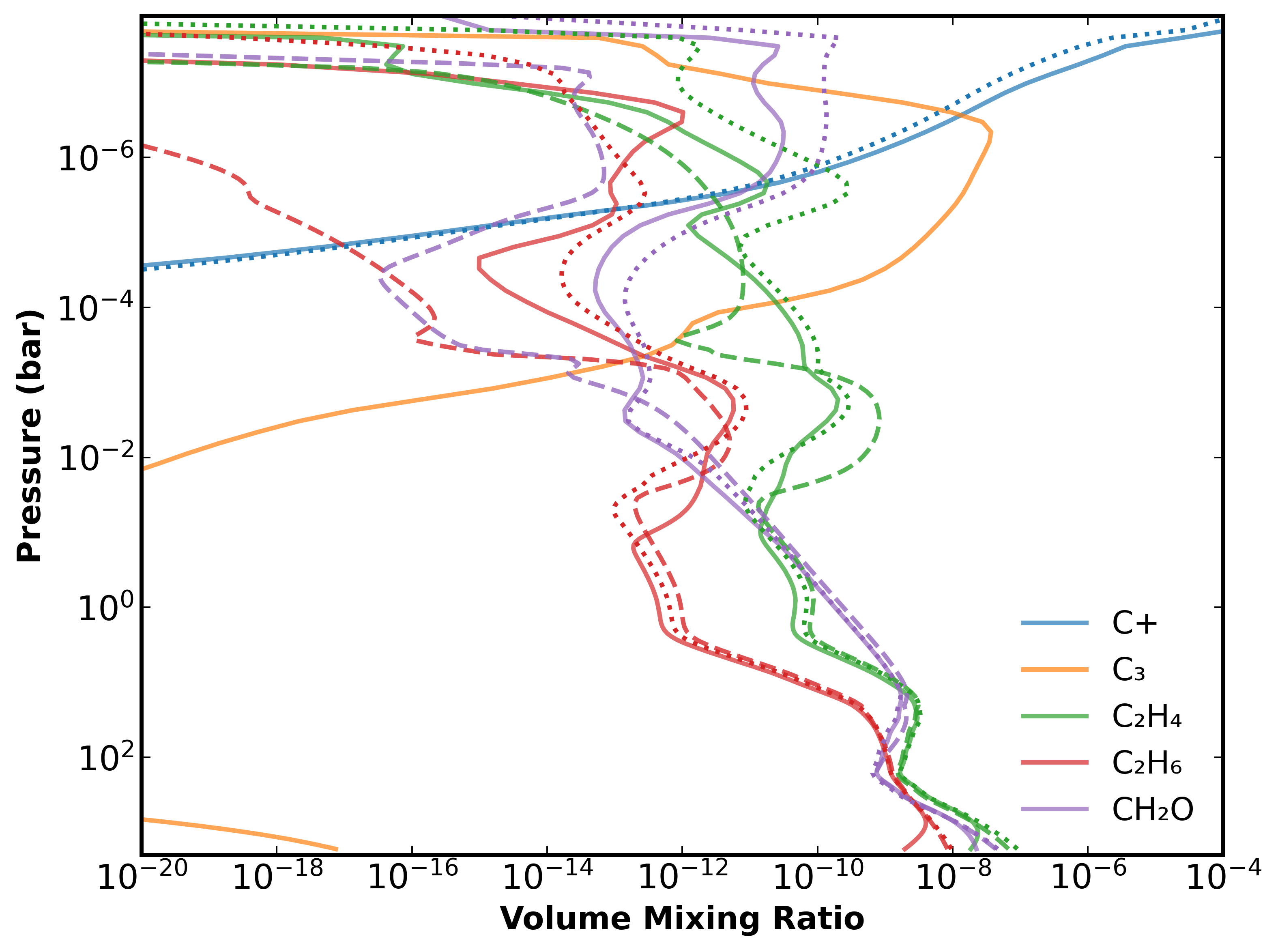}
		
		\caption{Volume mixing ratios of carbon-bearing species computed using three
			chemical networks with \texttt{XODIAC}: \texttt{XODIAC-2025} (solid lines),
			\texttt{XODIAC-STAND2020} (dotted lines), and \texttt{XODIAC-SNCHO} (dashed lines).}
		\label{fig:carbon_chemistry}
	\end{figure*}

	\begin{figure*}[p]
		\centering
		% First subplot
		% \begin{subfigure}[b]{0.45\textwidth}
			\centering
			\includegraphics[width=1.2\linewidth,angle=90, trim=2cm 0cm 2cm 0cm,clip]{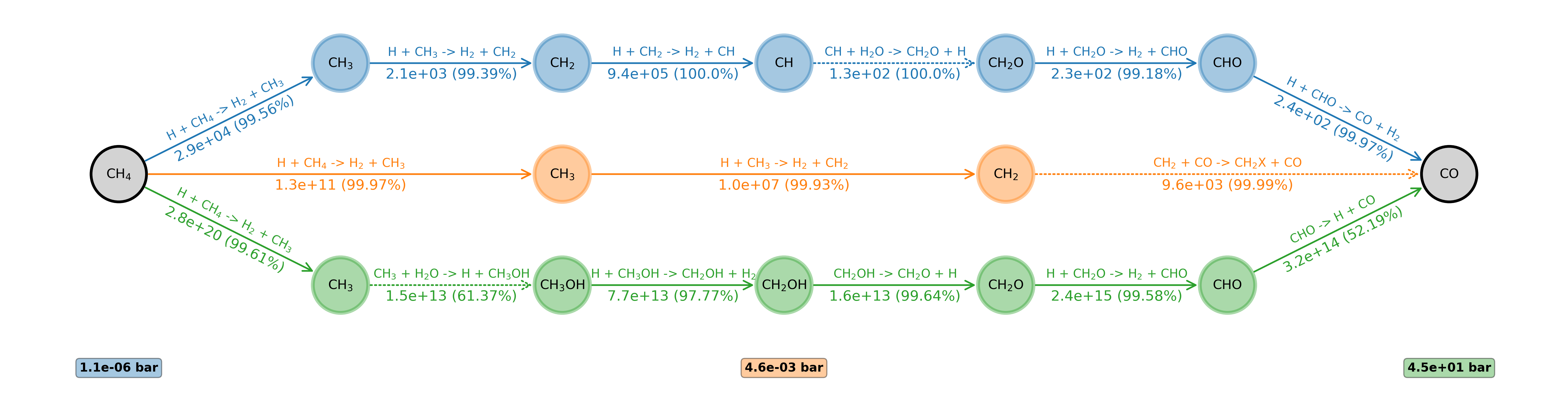}
			% \caption{}
			% \label{fig:CH4_to_CO}
			% \end{subfigure}
		% \hfill
		\hspace{2cm}
		% Second subplot
		% \begin{subfigure}[b]{0.45\textwidth}
			% \centering
			\includegraphics[width=1.2\linewidth,angle=90, trim=2cm 0cm 2cm 0cm,clip]{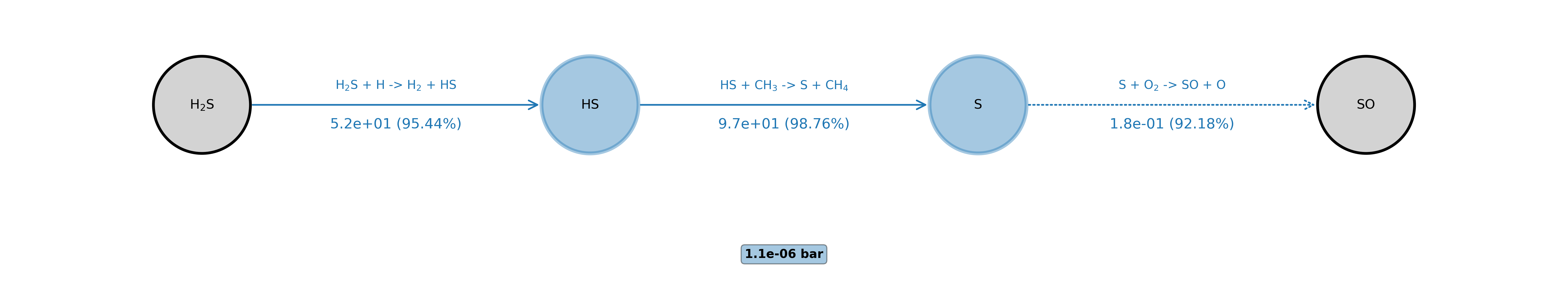}
			% \caption{}
			% \label{fig:H2S_to_SO}
			% \end{subfigure}
		
		% Main caption for the entire figure
		\caption{Chemical pathways for different species in various atmospheric layers. The dashed lines represent the rate-limiting step. \textbf{Left panel:} Dominant pathways from $\mathrm{CH_4}$ to $\mathrm{CO}$ in different atmospheric layers. \textbf{Right panel:} Dominant pathway to form $\mathrm{SO}$ from $\mathrm{H_2S}$ in the upper atmospheric layer.}
		\label{fig:reaction_pathway}
	\end{figure*}
	
	Recent \textit{JWST} observations have confirmed the presence of CO and CO$_2$ in the atmosphere of the exoplanet HD~189733~b \cite{fu2024hydrogensulfidemetalenrichedatmosphere}. These carbon-bearing species had been anticipated in earlier theoretical studies \cite{Mosses_2011, Tsai_2017, tsai2021comparative, moses2014chemical}. A comparative evaluation of CO and CO$_2$ chemistry across the \texttt{XODIAC-SNCHO}, \texttt{XODIAC-STAND2020}, and \texttt{XODIAC-2025} networks, along with other major (peak VMR $\ge 10^{-8}$) and minor ($10^{-14} \le \mathrm{VMR} < 10^{-8}$) carbon-bearing species, can provide deeper constraints on the global carbon chemistry in the atmosphere of HD~189733~b.
	
	CO remains the dominant carbon-bearing species throughout the atmosphere. However, in the deeper layers (pressures exceeding $\sim$10\,bar), CH$_4$ becomes the most abundant species, followed by C$_3$H$_7$ (present only in the \texttt{XODIAC-2025} network) and CO. With decreasing pressure, CO becomes the dominant species, surpassing CH$_4$ and CO$_2$, primarily due to quenching kinetics, as previously described by \citet{2011ApJ...738...72V}. To further investigate these chemical transitions, we identified the dominant reaction pathways responsible for CH$_4$–CO interconversion at three representative atmospheric layers (thermochemical, diffusive, and photochemical), as shown in the left panel of Figure \ref{fig:reaction_pathway}. In all cases, CH$_4$ is first converted to CH$_3$, after which the reaction pathways diverge depending on local environmental conditions. The rate-limiting step, defined as the slowest reaction in each pathway, varies with atmospheric altitude. Using the \texttt{XODIAC-2025} and \texttt{XODIAC-STAND2020} networks, we find significantly higher CH$_4$ abundances at pressures below $\sim$$10^{-3}$\,bar compared with \texttt{XODIAC-SNCHO}, while the CO mixing ratio remains essentially unchanged. In contrast, CO$_2$ abundances are markedly lower in \texttt{XODIAC-2025} and \texttt{XODIAC-STAND2020}, which accounts for the relative increase in CH$_4$.
	
	We also examined CO$_2$ production pathways at high altitudes. Across all networks, CH$_4$ abundances decrease sharply below $\sim$$10^{-3}$\,bar, while CO remains stable. Consequently, tracking the redistribution of carbon from CH$_4$ becomes essential. As shown in Figure \ref{fig:carbon_chemistry}, CO$_2$ abundances increase in tandem with CH$_4$ depletion at higher altitudes for all networks. A detailed analysis of the dominant CH$_4$–CO$_2$ pathways reveals CO as the final precursor, with identical pathways and rate-limiting steps to those in CH$_4$–CO interconversion. Thus, only the final CO $\rightarrow$ CO$_2$ step is explicitly shown:
	
	\begin{equation}
		\mathrm{CO} + \mathrm{OH} \longrightarrow \mathrm{CO_2} + \mathrm{H}
	\end{equation}
	
	Using the \texttt{XODIAC-2025} network, we find that near the surface, where thermochemistry dominates, in addition to \ce{CH4} and \ce{CO}, another notable species, \ce{C3H7}, reaches a significant abundance (VMR $\sim 10^{-4}$). Its abundance decreases progressively with altitude, falling below $10^{-14}$ at pressures around $10^{-1}$\,bar. The primary destruction pathway for C$_3$H$_7$ is:
	
	\begin{equation} \label{C3H7_des}
		\mathrm{H} + \mathrm{C_3H_7} \longrightarrow \mathrm{C_3H_6} + \mathrm{H_2}
	\end{equation}
	
	CH$_3$ shows a similar initial decline in abundance, decreasing toward $10^{-1}$\,bar before increasing again at higher altitudes, and stabilizing between VMR values of $10^{-7}$ and $10^{-9}$. In the middle atmosphere (around $10^{-3}$\,bar), its chemistry is dominated by the neutral–neutral reaction and its reverse:
	
	\begin{equation}
		\mathrm{H} + \mathrm{CH_4} \leftrightharpoons \mathrm{H_2} + \mathrm{CH_3}
	\end{equation}
	
	The third most abundant hydrocarbon, C$_2$H$_2$, maintains a relatively steady mixing ratio ranging from $10^{-8}$ to $10^{-11}$, exhibiting a vertical trend similar to that of CH$_3$ from the surface up to $\sim$$10^{-3}$ bar. The dominant formation and destruction processes for C$_2$H$_2$ involve:
	
	\begin{equation} 
		\mathrm{C_2H_3} + \mathrm{M} \leftrightharpoons \mathrm{H} + \mathrm{C_2H_2} + \mathrm{M}
	\end{equation}
	
	In the uppermost atmospheric layers (pressures $<$ $10^{-6}$ bar), all heavier carbon-bearing species (CO, CH$_4$, CO$_2$, C$_3$H$_7$, C$_2$H$_2$, CH$_3$, C$_2$H$_4$, and C$_3$) decline sharply, giving way to lighter, ionized, or electronically excited species. At these altitudes, atomic carbon (C) and ionized carbon (C$^+$) dominate, underscoring the pivotal role of photochemistry under intense stellar irradiation. Key reactions include:
	
	\begin{equation}
		\mathrm{CH} + \mathrm{H} \longrightarrow \mathrm{C} + \mathrm{H_2}
	\end{equation}
	\begin{equation}
		\ce{CO ->[\text{h$\nu$}] C + O}
	\end{equation}
	\begin{equation}
		\ce{C ->[\text{h$\nu$}] C$^+$ + e$^-$}
	\end{equation}
	
	In summary, all three networks indicate that in the deep atmosphere ($\sim$500~bar), \ce{CH4} is the most abundant species, followed by \ce{CO}. In \texttt{XODIAC-2025}, the third most abundant species is \ce{C3H7}, whereas in \texttt{XODIAC-STAND2020} and \texttt{XODIAC-SNCHO} it is the \ce{CH3} radical. In the middle atmosphere ($\sim10^{-2}$~bar), \ce{CO} becomes the dominant species across all networks, followed by \ce{CH4} and \ce{CO2}.

	\subsection{Comparative Sulfur Chemistry}
	
	In the atmosphere of HD 189733 b, $\mathrm{H_2S}$ has recently been observed with JWST \citep{fu2024hydrogensulfidemetalenrichedatmosphere}. In the lower atmosphere, from pressures of $10^{3}$\,bar to $10^{-1}$\,bar, $\mathrm{H_2S}$ is the dominant sulfur carrier. Its primary formation pathway is:
	
	\begin{equation} \label{eq: H2S formation}
		\mathrm{HS} + \mathrm{H_2} \longrightarrow \mathrm{H} + \mathrm{H_2S}
	\end{equation}
	
	In the upper atmosphere, however, $\mathrm{H_2S}$ is destroyed by reactions with $\mathrm{H}$ radicals, produced via photochemistry, through:
	
	\begin{equation} \label{eq: H2S destruction}
		\mathrm{H_2S} + \mathrm{H} \longrightarrow \mathrm{HS} + \mathrm{H_2}
	\end{equation}
	
	High pressures favor the conversion of $\mathrm{HS}$ into $\mathrm{H_2S}$, consistent with the trends reported by \citet{Hobbs_2021}. Depending on pressure, temperature, and other atmospheric conditions, $\mathrm{HS}$ acts as an intermediate radical that can be converted either into $\mathrm{H_2S}$ or atomic sulfur (S). It is also a crucial intermediate in the formation of many other sulfur-bearing species.
	
	\begin{figure*}[ht]
		\centering
		
		\includegraphics[width=0.48\textwidth]{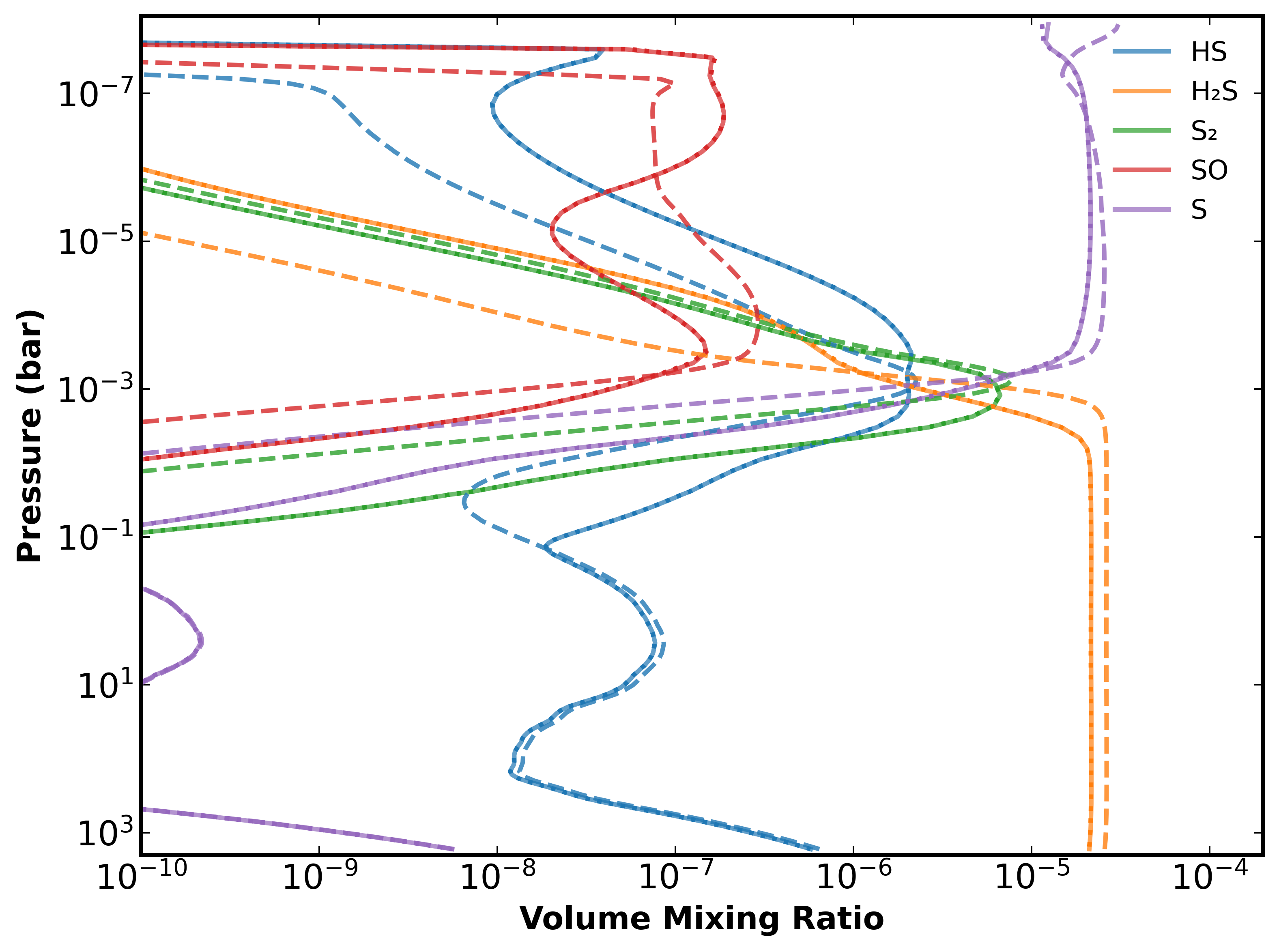}
		\hfill
		\includegraphics[width=0.48\textwidth]{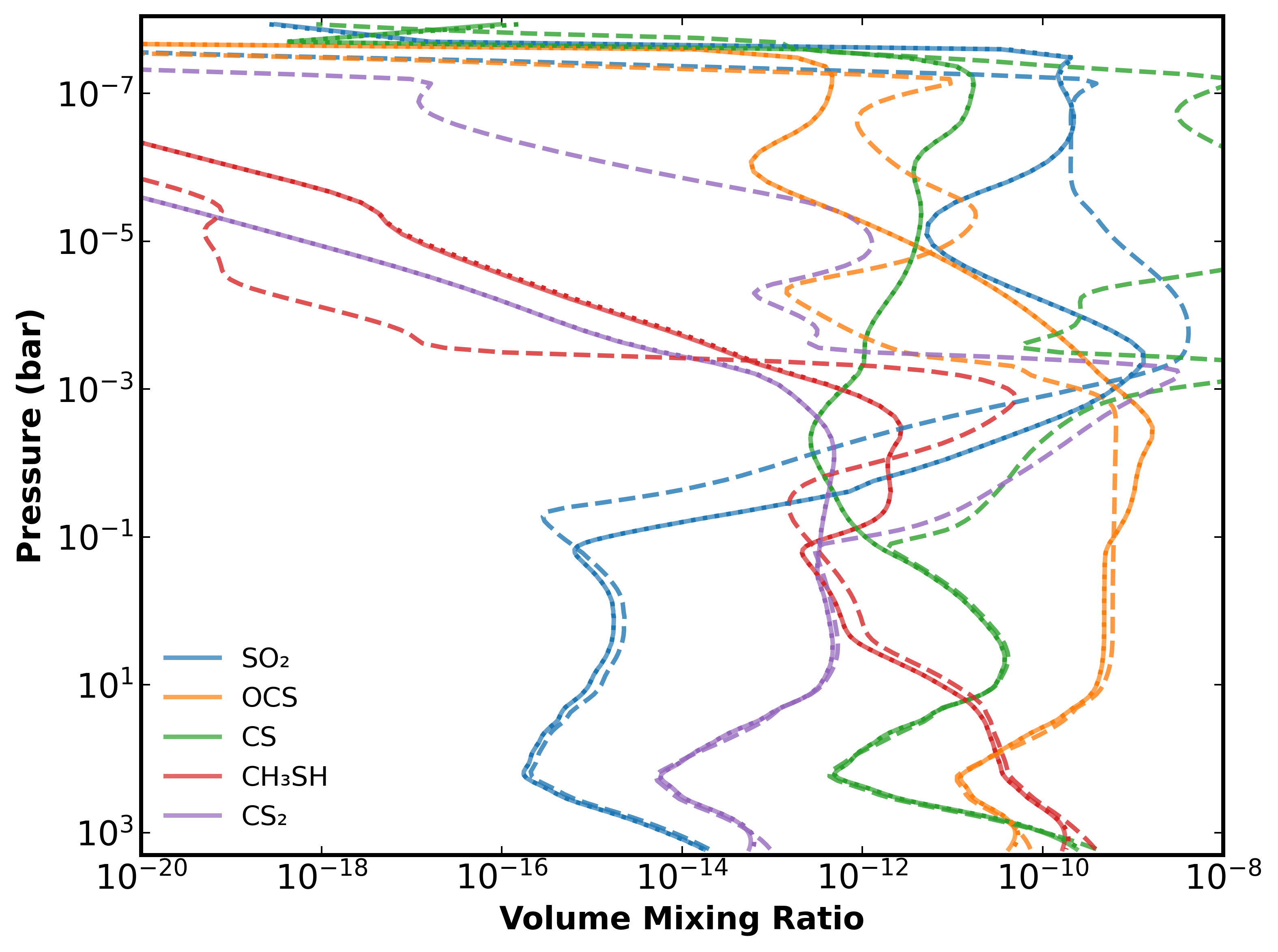}
		
		\caption{Volume mixing ratios of sulfur-bearing species computed using three
			chemical networks with \texttt{XODIAC}: \texttt{XODIAC-2025} (solid lines),
			\texttt{XODIAC-STAND2020} (dotted lines), and \texttt{XODIAC-SNCHO} (dashed lines).}
		\label{fig:sulfur_chemistry}
	\end{figure*}

	Figure~\ref{fig:sulfur_chemistry} shows the major sulfur species (peak VMR $>10^{-8}$) in the left panel. In the lower atmosphere, up to $10^{-3}$\,bar, the abundance of H$_2$S remains quenched. Above this level, H$_2$S begins to lose its dominance over other sulfur species. Notably, below $10^{-3}$\,bar, H$_2$S and SO show much larger abundance variations compared to other major sulfur molecules, indicating that H$_2$S is converted to SO. 
	% Higher in the atmosphere, photolysis of H$_2$O
	% \begin{equation}
		% \ce{H2O ->[\text{h$\nu$}] OH + H}
		% \end{equation}
	% \noindent makes OH a key agent for H$_2$S-to-SO conversion. 
	The overall H$_2$S-to-SO conversion pathway is illustrated in the right panel of Figure~\ref{fig:reaction_pathway}. 
	
	Between $10^{-1}$\,bar and $10^{-3}$\,bar, $\mathrm{S_2}$ becomes the dominant sulfur carrier as $\mathrm{H_2S}$ starts to dissociate, but insufficient atomic S has yet formed to dominate. The main $\mathrm{S_2}$-forming reaction, solely dominating, is:
	
	\begin{equation} \label{eq: S2 major reaction}
		\mathrm{HS} + \mathrm{S} \longrightarrow \mathrm{S_2} + \mathrm{H}
	\end{equation}
	
	Figure~\ref{fig:sulfur_chemistry} also shows that atomic S dominates from $10^{-3}$\,bar to $10^{-7}$\,bar. The primary formation pathway is:
	
	\begin{equation} \label{eq: S forming reaction}
		\mathrm{H} + \mathrm{HS} \longrightarrow \mathrm{S} + \mathrm{H_2}
	\end{equation}
	
	High temperatures favor $\mathrm{H_2S}$ dissociation into $\mathrm{H_2}$ and S, leading to S dominance in hotter regions. In the upper atmosphere (low pressure, high temperature), the increase in S abundance is further driven by enhanced HS levels. Reaction~\eqref{eq: S forming reaction} accounts for 98\% of atomic S production there. 
	
	Other sulfur species, such as SO and $\mathrm{SO_2}$, also increase in abundance with rising temperature and decreasing pressure. SO is primarily formed from the oxidation of atomic S, which is abundant in the upper atmosphere, via:
	\begin{equation} \label{eq:SO_formation}
		\mathrm{S} + \mathrm{OH} \longrightarrow \mathrm{SO} + \mathrm{H}
	\end{equation}
	
	\noindent and remains a major sulfur species throughout much of the atmosphere. Below $10^{-5}$\,bar, SO becomes the second most abundant sulfur carrier (Figure~\ref{fig:sulfur_chemistry}).
	
	Sulfur dioxide ($\mathrm{SO_2}$) shows enhanced abundances between $10^{-3}$\,bar and $10^{-6}$\,bar. In the range $10^{-3}$\,bar to $10^{-4}$\,bar, the dominant $\mathrm{SO_2}$ formation pathways, contributing roughly equally, are:
	\begin{equation} \label{eq:SO2_from_SO}
		2\,\mathrm{SO} \longrightarrow \mathrm{SO_2} + \mathrm{S}
	\end{equation}
	\begin{equation} \label{eq:SO2_from_OH}
		\mathrm{SO} + \mathrm{OH} \longrightarrow \mathrm{SO_2} + \mathrm{H}
	\end{equation}
	At pressures below $10^{-4}$\,bar, down to approximately $10^{-7}$\,bar, reaction~\eqref{eq:SO2_from_OH} becomes the dominant $\mathrm{SO_2}$ formation pathway.

	CS, a minor carbon-bearing species, is relatively stable due to the strong C$\equiv$S triple bond, making it less reactive with other atmospheric species. Its mixing ratio remains in the range $\sim$$10^{-13}$--$10^{-9}$ throughout the atmosphere, up to $10^{-7}$\,bar. The main CS-forming reactions are:
	
	\begin{equation} \label{lower layer}
		\mathrm{OCS} + \mathrm{H} \longrightarrow \mathrm{CS} + \mathrm{OH}
	\end{equation}
	\begin{equation} \label{middle layer}
		\mathrm{CS_2} + \mathrm{S} \longrightarrow \mathrm{CS} + \mathrm{S_2}
	\end{equation}
	\begin{equation} \label{upper layer}
		\mathrm{OCS} + \mathrm{C} \longrightarrow \mathrm{CS} + \mathrm{CO}
	\end{equation}
	\begin{equation} \label{upper_layer}
		\ce{OCS ->[h\nu] CS + O(^1D)}
	\end{equation}
	
	Different reactions dominate CS formation at different pressures: reaction~\eqref{lower layer} from $10^{3}$\,bar to $10^{-2}$\,bar, reaction~\eqref{middle layer} between $10^{-2}$\,bar and $10^{-3}$\,bar, and reaction~\eqref{upper layer} below $10^{-4}$\,bar to $10^{-5}$\,bar, where it contributes to the formation of CS with a share of $> 70$\%.
	In the pressure range from $10^{-5}$ bar to the upper atmosphere, \ce{CS} is produced through both photodissociation processes and neutral–neutral reactions, as indicated by reactions \ref{upper_layer} and \ref{upper layer}.
	
	OCS shows greater variability than CS because its C=O double bond is easier to break than the C$\equiv$S triple bond. OCS is produced mainly via:
	
	\begin{equation} \label{upper}
		\mathrm{HS} + \mathrm{CO} \longrightarrow \mathrm{OCS} + \mathrm{H}
	\end{equation}
	\begin{equation} \label{middle}
		\mathrm{CO} + \mathrm{S} + \mathrm{M} \longrightarrow \mathrm{OCS} + \mathrm{M}
	\end{equation}
	\begin{equation} \label{lower}
		\mathrm{CS} + \mathrm{OH} \longrightarrow \mathrm{OCS} + \mathrm{H}
	\end{equation}
	
	From $10^{3}$\,bar to 10\,bar, reaction~\eqref{upper} dominates with a 50--$\mathrm{\sim80}$\% share, followed by reaction~\eqref{middle} at 25--50\%. Between 10\,bar and $10^{-1}$\,bar, reactions~\ref{middle} and~\ref{upper} constitute the dominant pathways for the formation of \ce{OCS}.
	% reaction~\eqref{middle} leads (50\%), with reaction~\eqref{upper} a close second (48\%). 
	From $10^{-1}$\,bar to $10^{-6}$\,bar, these two pathways alternate in dominance due to similar contributions. At the highest altitudes ($\sim$$10^{-6}$\,bar), reaction~\eqref{lower} overwhelmingly dominates OCS formation, accounting for $\sim$97\%.
	
	Throughout the atmosphere, the primary formation pathway of \ce{CH3SH} is given by reaction~\ref{CH3SH major forming}. At the highest altitudes (pressures $<10^{-7}$~bar), reaction~\ref{CH3SH major forming above 1e-8} becomes the dominant source of \ce{CH3SH}.
	
	\begin{equation} \label{CH3SH major forming}
		\mathrm{CH_3} + \mathrm{HS} + \mathrm{M} \longrightarrow \mathrm{CH_3SH} + \mathrm{M}
	\end{equation}
	
	\begin{equation} \label{CH3SH major forming above 1e-8}
		\mathrm{CH_3} + \mathrm{H_2S} \longrightarrow \mathrm{CH_3SH} + \mathrm{H}
	\end{equation}
	
	Overall, the abundance trends of major species predicted by the \texttt{XODIAC-2025} network show strong agreement with those obtained from the \texttt{XODIAC-STAND2020} network. In contrast, \texttt{XODIAC-SNCHO} shows only a minor deviation of about one order of magnitude in the pressure range $10^{-1}$ to $10^{-6}$~bar. For example, $\mathrm{SO_2}$ undergoes greater dissociation in the \texttt{XODIAC-2025} network due to reactions with both $\mathrm{S}$ and $\mathrm{H}$, which are absent in \texttt{XODIAC-SNCHO}.

	\begin{equation} \label{SO2 major DESTRUCTION}
		\mathrm{S} + \mathrm{SO_2} \longrightarrow 2\,\mathrm{SO}
	\end{equation}
	
	\begin{equation} \label{so2 major destruction}
		\mathrm{H} + \mathrm{SO_2} \longrightarrow \mathrm{SO} + \mathrm{OH}
	\end{equation}
	
	A detailed analysis of the major destruction pathways of \(\mathrm{SO_2}\) in the pressure range between \(10^{-1}\) and \(10^{-3}\)~bar reveals that, in the \texttt{XODIAC-2025} network, reactions~\eqref{SO2 major DESTRUCTION} and~\eqref{so2 major destruction} are the primary contributors. In contrast, for the \texttt{XODIAC-SNCHO} network, reaction~\eqref{so2 major destruction} dominates in this region. At higher altitudes, around \(10^{-6}\)~bar, the observed deviations arise due to photodissociation processes. In the \texttt{XODIAC-2025} network, \(\mathrm{SO_2}\) destruction is dominated by photodissociation from approximately \(10^{-4}\)~bar up to \(\sim10^{-7}\)~bar, whereas in the \texttt{XODIAC-SNCHO} network, photoreactions begin to dominate at slightly higher altitudes, around \(10^{-5}\)~bar. These subtle differences in the altitude range where photochemical processes become dominant contribute to the minor discrepancies observed in the \(\mathrm{SO_2}\) abundance profiles between the two networks. The major photodissociation reactions for \(\mathrm{SO_2}\) are:
	
	\begin{equation} \label{SO2 photoreaction}
		\ce{SO_2 ->[\text{h$\nu$}] SO + O}
	\end{equation}
	
	\begin{equation} \label{SO2 major photoreaction}
		\ce{SO_2 ->[\text{h$\nu$}] S + O_2}
	\end{equation}
	
	Similar to \(\mathrm{SO_2}\), the minor sulfur-bearing and photochemically active species \(\mathrm{CS_2}\) shows a similar abundance profile throughout the atmosphere when comparing the \texttt{XODIAC-2025} and \texttt{XODIAC-STAND2020} networks. The \texttt{XODIAC-2025} network also shows good agreement with \texttt{XODIAC-SNCHO} up to a pressure of \(10^{-1}\)~bar for $\mathrm{CS_2}$ abundances. However, at lower pressures toward the upper atmosphere, noticeable deviations appear.
	
	From pressures around $10^{-1}$~bar to $10^{-3}$~bar, the major destruction reaction that dominates solely in the \texttt{XODIAC-2025} network is represented by reaction~\ref{middle layer}.
	% \begin{equation} \label{CS2 major DESTRUCTION 1e-1 to 1e-3}
		% \mathrm{S} + \mathrm{CS_2} \longrightarrow \mathrm{CS} + \mathrm{S_2}
		% \end{equation}
	
	In contrast, in the \texttt{XODIAC-SNCHO} network, $\mathrm{CS_2}$ is primarily destroyed by $\mathrm{CO}$ (70\%–75\%) via reaction~\eqref{CS2 major DESTRUCTION 1e-1 to 1e-3 vulcan major} and secondarily (25\%–30\%) by $\mathrm{H}$ via reaction~\eqref{CS2 major DESTRUCTION 1e-1 to 1e-3 vulcan minor}.
	
	\begin{equation} \label{CS2 major DESTRUCTION 1e-1 to 1e-3 vulcan major}
		\mathrm{CO} + \mathrm{CS_2} \longrightarrow \mathrm{CS} + \mathrm{OCS}
	\end{equation}
	
	\begin{equation} \label{CS2 major DESTRUCTION 1e-1 to 1e-3 vulcan minor}
		\mathrm{H} + \mathrm{CS_2} \longrightarrow \mathrm{CS} + \mathrm{HS}
	\end{equation}
	
	As with $\mathrm{SO_2}$, from $10^{-3}$~bar to the upper atmosphere, photodissociation plays a key role in driving deviations. In the \texttt{XODIAC-2025} network, between $10^{-3}$~bar and $10^{-4}$~bar, $\mathrm{CS_2}$ destruction is still primarily governed by reaction~\ref{middle layer}, with minor contributions from photodissociation. At higher altitudes ($10^{-4}$~bar to $10^{-7}$~bar), photodissociation 
	% (reaction~\eqref{CS2 photo dissociation xodiac}) 
	becomes the sole dominant pathway, via
	
	\begin{equation} \label{CS2 photo dissociation xodiac}
		\ce{CS_2 ->[\text{h$\nu$}] CS + S}
	\end{equation}
	
	In the \texttt{XODIAC-SNCHO} network, the major destruction pathways for \(\mathrm{CS_2}\) are reactions~\eqref{CS2 major DESTRUCTION 1e-1 to 1e-3 vulcan major} and~\eqref{CS2 major DESTRUCTION 1e-1 to 1e-3 vulcan minor}. Between \(10^{-3}\) and \(\sim10^{-4}\)~bar, photodissociation (reaction~\eqref{CS2 photo dissociation xodiac}) emerges as a minor pathway, while reaction~\eqref{CS2 major DESTRUCTION 1e-1 to 1e-3 vulcan minor} remains dominant. Between \(10^{-4}\) and \(10^{-5}\)~bar, photodissociation becomes a significant contributor, sharing dominance with reaction~\eqref{CS2 major DESTRUCTION 1e-1 to 1e-3 vulcan minor}. Below \(10^{-5}\)~bar, photodissociation becomes the primary destruction route. The slower $\mathrm{CS_2}$ destruction rate in the \texttt{XODIAC-SNCHO} network, particularly in the $10^{-1}$ to $\sim$$10^{-4}$~bar range, results from the weaker contribution of neutral–neutral reactions compared to \texttt{XODIAC-2025}. This difference extends into the upper atmosphere, where photodissociative processes further amplify deviations in abundance profiles, similar to the case of $\mathrm{SO_2}$.
	
	In summary, sulfur chemistry is broadly consistent across the three networks. In the lower atmosphere ($\sim$500~bar), \ce{H2S} is the most abundant sulfur-bearing species, followed by \ce{HS}. In \texttt{XODIAC-2025}, \ce{S} is present but with a mixing ratio below $10^{-8}$, whereas in the other two networks it exceeds this threshold. In the middle atmosphere ($\sim10^{-2}$~bar), \ce{H2S} and \ce{HS} remain dominant, with \ce{S2} rapidly varying, coming very close to \ce{HS} in abundance. In the upper atmosphere ($\sim10^{-6}$~bar), all networks predict atomic sulfur (\ce{S}) as the most abundant species, followed by sulfur monoxide (\ce{SO}) and \ce{HS}.

	\subsection{Comparative Phosphorus Chemistry}
	
\begin{figure*}[ht]
		\centering
		
		\includegraphics[width=0.48\textwidth]{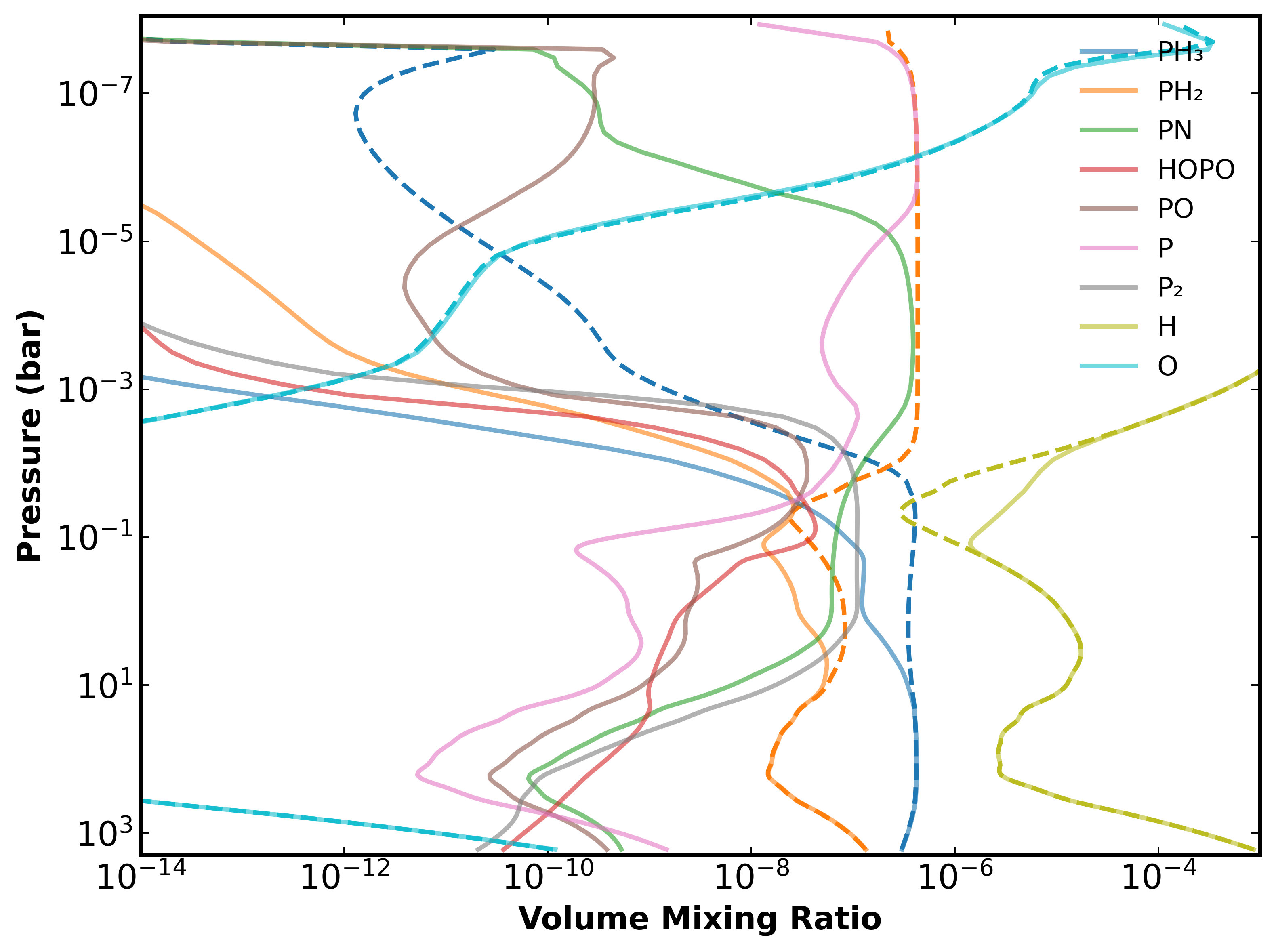}
		\hfill
		\includegraphics[width=0.48\textwidth]{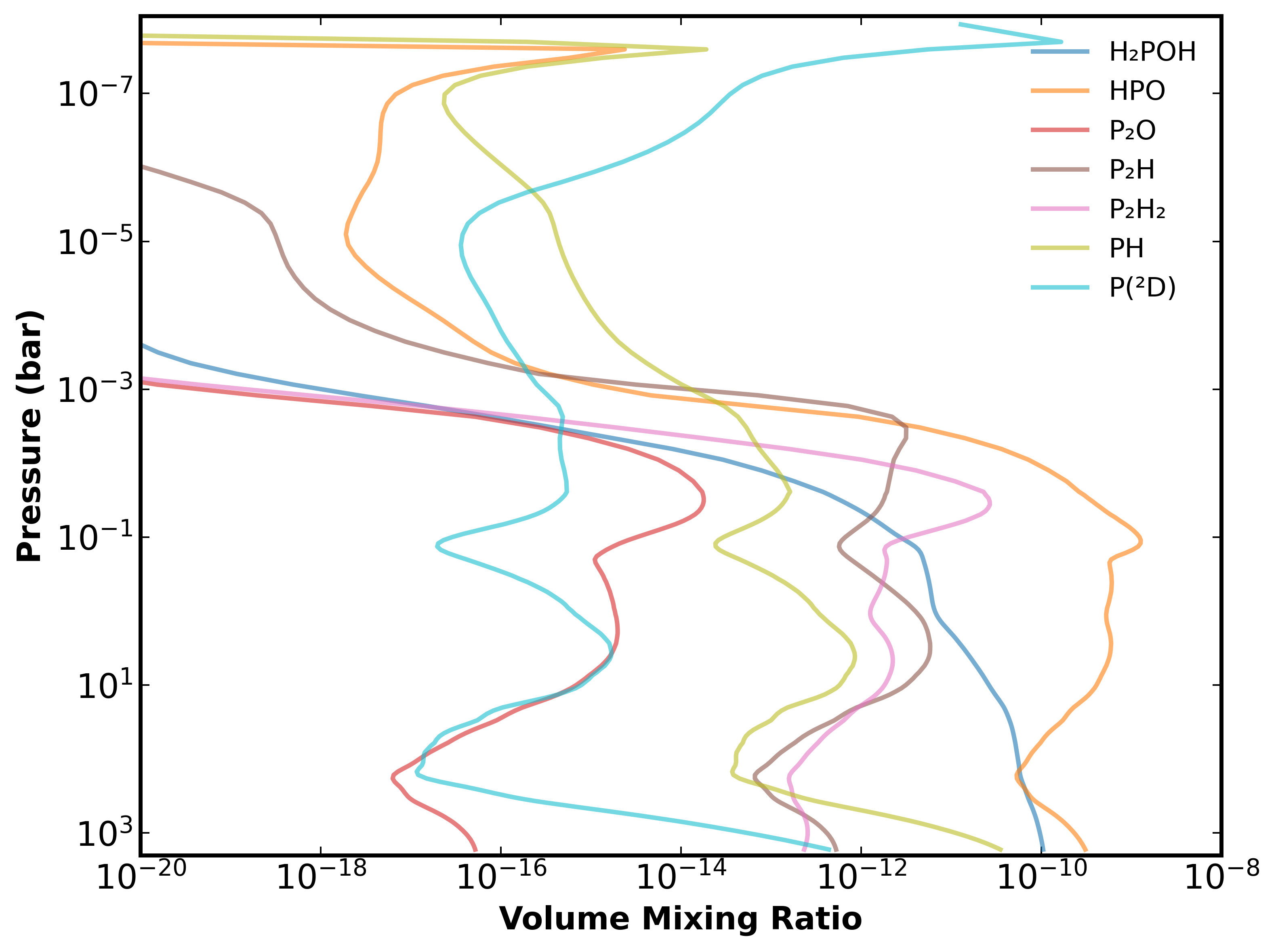}
		
		\caption{Volume mixing ratios of phosphorus-bearing species computed using two
			chemical networks with \texttt{XODIAC}: \texttt{XODIAC-2025} (solid lines) and
			\texttt{XODIAC-STAND2020} (dashed lines). Phosphorus chemistry is not included
			in the \texttt{XODIAC-SNCHO} network.}
		\label{fig:phosphorus_chemistry}
	\end{figure*}

	For the hot Jupiter HD 189733 b, Figure \ref{fig:phosphorus_chemistry} shows the VMR profiles of phosphorus-bearing molecules computed using the \texttt{XODIAC-2025} and \texttt{XODIAC-STAND2020} chemical networks. In \texttt{XODIAC-STAND2020}, phosphorus chemistry is limited to only two species, PH$_3$ and PH$_2$. In contrast, \texttt{XODIAC-2025} incorporates major updates to the phosphorus network, introducing 24 phosphorus species and leading to significant deviations in the predicted VMRs of PH$_3$ and PH$_2$ (Figure~\ref{fig:phosphorus_chemistry}).

	Based on the \texttt{XODIAC-2025} results, we identify PH$_3$, P$_2$, PN, and atomic P as the dominant phosphorus-bearing species at different atmospheric altitudes. 
		In the deepest layers (up to 10$^{-1}$~bar), where thermochemical equilibrium prevails, PH$_3$ is the most abundant phosphorus carrier, followed by P$_2$ and PN. In this region, PH$_3$ remains close to equilibrium, with nearly equal forward and reverse reaction rates (Figure~\ref{fig:phosphorus_chemistry}).

	% In the deepest layers ($>$ 220 bar), where thermochemistry dominates, P$_2$O$_2$ is the most abundant form of phosphorus, followed by PH$_3$ and PH$_2$. The main reactions contributing to P$_2$O$_2$ formation and destruction are:
	
	% \begin{equation}
		% \mathrm{OH} + \mathrm{P_2O} \leftrightharpoons \mathrm{H} + \mathrm{P_2O_2}
		% \end{equation}
	% \begin{equation}
		% \mathrm{PO} + \mathrm{HPO} \leftrightharpoons \mathrm{H} + \mathrm{P_2O_2}
		% \end{equation}
	% \begin{equation}
		% \mathrm{PO} + \mathrm{P_2O} \leftrightharpoons \mathrm{P} + \mathrm{P_2O_2}
		% \end{equation}
	% \begin{equation}
		% \mathrm{P_2} + \mathrm{P_2O_2} \longrightarrow 2\mathrm{P_2O}
		% \end{equation}
	
	% Above this region, from $\sim$100 bar to $\sim$1 bar, PH$_3$ becomes the dominant phosphorus-bearing species. Here, thermochemistry alone governs its abundance, with PH$_3$ being quenched between $\sim$200 bar and $\sim$15 bar, where forward and reverse reaction rates are nearly equal (Figure \ref{fig:phosphorus_chemistry}):
	
	\begin{equation} \label{eq: PH3-PH2}
		\mathrm{PH_3} + \mathrm{H} \leftrightharpoons \mathrm{PH_2} + \mathrm{H_2}
	\end{equation}
	
	In the \texttt{XODIAC-STAND2020} network, PH$_3$ dominates until the middle atmosphere ($\sim$10$^{-2}$ bar) through the reverse pathway of reaction~\ref{eq: PH3-PH2}. At higher altitudes, the concentration of H radicals increases (Figure \ref{fig:phosphorus_chemistry}), favoring the forward pathway of reaction~\ref{eq: PH3-PH2}. Since PH$_2$ is the only lighter phosphorus species besides PH$_3$, it dominates from $\sim$10$^{-2}$ bar to the top of the atmosphere.
	
	In contrast, in the \texttt{XODIAC-2025} network, P$_2$ serves as the major phosphorus reservoir between $\sim$10$^{-1}$~bar and $\sim$10$^{-2}$~bar, with a VMR of $\sim$10$^{-7}$, and remains stable through a thermodynamic three-body reaction showing nearly equal forward and reverse rates:

	\begin{equation}\label{P2_H_M}
		\mathrm{P_2} + \mathrm{H} + \mathrm{M} \leftrightharpoons \mathrm{P_2H} + \mathrm{M}
	\end{equation}
	
	Between $\sim$10$^{-2}$ bar and $\sim$10$^{-5}$ bar, PN serves as the dominant phosphorus reservoir. The key PN-forming reactions are:
	
	\begin{equation} \label{PO+HN}
		\mathrm{PO} + \mathrm{HN} \longrightarrow \mathrm{PN} + \mathrm{OH}
	\end{equation}
	\begin{equation} \label{P+HN}
		\mathrm{P} + \mathrm{HN} \longrightarrow \mathrm{PN} + \mathrm{H}
	\end{equation}
	\begin{equation} \label{PO+N}
		\mathrm{PO} + \mathrm{N} \longrightarrow \mathrm{PN} + \mathrm{O}
	\end{equation}
	
	Reaction~\eqref{PO+HN} dominates from $\sim$500 bar to $\sim$10$^{-2}$ bar, reaction~\eqref{P+HN} from $10^{-2}$ to $10^{-6}$ bar, and reaction~\eqref{PO+N} below $10^{-6}$ bar. Owing to the strong triple bond between P and N, PN is highly resistant to photodissociation, making it the principal phosphorus-bearing species in the mid-atmosphere. 
	Around 10$^{-2}$ bar, PN oxidation leads to dissociation of the PN molecule through:
	\begin{equation} \label{PN_OH}
		\mathrm{PN} + \mathrm{OH} \longrightarrow \mathrm{PO} + \mathrm{HN}
	\end{equation}
	\begin{equation}\label{PN_O}
		\mathrm{PN} + \mathrm{O} \longrightarrow \mathrm{PO} + \mathrm{N}
	\end{equation}
	
	However, between 10$^{-2}$ bar and 10$^{-5}$ bar, the major PN destruction proceeds through:
	\begin{equation}\label{PN_N}
		\mathrm{PN} + \mathrm{N} \longrightarrow \mathrm{P} + \mathrm{N_2}
	\end{equation}
	
	Below $\sim$10$^{-5}$ bar, atomic P becomes the most abundant phosphorus species as PN converts to P via
	\begin{equation} \label{PN and C}
		\mathrm{PN} + \mathrm{C} \longrightarrow \mathrm{P} + \mathrm{CN}
	\end{equation}
	\begin{equation}\label{PO to P + O}
		\ce{\mathrm{PO} ->[\text{h$\nu$}] \mathrm{P} + \mathrm{O}}
	\end{equation}
	
	HOPO also appears with a mixing ratio of $\sim$10$^{-8}$ in the mid-atmosphere. Its primary formation pathways occur at $\sim$10$^3$–10$^{-1}$ bar, $\sim$10$^{-3}$–10$^{-4}$ bar, and below $\sim$10$^{-7}$ bar through:
	\begin{equation} \label{PO to HOPO}
		\mathrm{PO} + \mathrm{H_2O} \longrightarrow \mathrm{HOPO} + \mathrm{H}
	\end{equation}
	
	In other pressure regimes, the dominant pathway is:
	\begin{equation} \label{PO2 to HOPO}
		\mathrm{PO_2} + \mathrm{H_2} \longrightarrow \mathrm{HOPO} + \mathrm{H}
	\end{equation}

	Below $\sim$10$^{-2}$ bar, photochemistry becomes important. Increasing concentrations of H, O, OH, and P radicals drive the destruction of large phosphorus molecules into smaller species, reducing the abundances of HOPO, PH$_3$, PH$_2$, and P$_2$:
	
	\begin{equation}
		\ce{\mathrm{HOPO} ->[\text{h$\nu$}] \mathrm{H} + \mathrm{PO_2}}
	\end{equation}
	\begin{equation} \label{HOPO to PO}
		\mathrm{HOPO} + \mathrm{H} \longrightarrow \mathrm{PO} + \mathrm{H_2O}
	\end{equation}
	\begin{equation}
		\mathrm{PH_3} + \mathrm{H} \longrightarrow \mathrm{PH_2} + \mathrm{H_2}
	\end{equation}
	\begin{equation}
		\mathrm{PH_2} + \mathrm{H} \longrightarrow \mathrm{PH} + \mathrm{H_2}
	\end{equation}
	\begin{equation}
		\mathrm{P_2} + \mathrm{H} + \mathrm{M} \longrightarrow \mathrm{P_2H} + \mathrm{M}
	\end{equation}
	\begin{equation} \label{P2_photo}
		\ce{\mathrm{P_2} ->[\text{h$\nu$}] \mathrm{P} + \mathrm{P}}
	\end{equation}
	
	% From $\sim10^{-3}$ bar and below, the reaction~\ref{P2_photo} starts becoming a major photochemical pathway for the production of P atoms. 
	
	From $\sim$10$^{-2}$ to $10^{-3}$ bar, reaction~\ref{HOPO to PO} contributes most to PO formation, while at higher altitudes reaction~\ref{P to PO} dominates ($>$90\% contribution):
	
	\begin{equation} \label{P to PO}
		\mathrm{P} + \mathrm{OH} \longrightarrow \mathrm{PO} + \mathrm{H}
	\end{equation}
	
	Atomic phosphorus formation primarily results from the destruction reaction~\ref{PN and C} of PN in the upper atmosphere. Between $\sim$10$^3$ bar and 10$^{-4}$ bar, the reaction:
	
	\begin{equation} \label{P_2D to P}
		\mathrm{P({}^2D)} + \mathrm{H_2} \longrightarrow \mathrm{P} + \mathrm{H_2}
	\end{equation}
	
	plays a key role, followed by reaction~\ref{PN and C}. In the highest layers ($<$10$^{-7}$ bar), where photochemical activity is intense, reaction~\ref{PO to P + O} becomes the dominant pathway for both PO destruction and atomic phosphorus production.

	% \begin{equation} \label{PN and C}
		% \mathrm{PN} + \mathrm{C} \longrightarrow \mathrm{P} + \mathrm{CN}
		% \end{equation}
	% \begin{equation} \label{PO to P + O}
		% \ce{\mathrm{PO} ->[\text{h$\nu$}] \mathrm{P} + \mathrm{O}}
		% \end{equation}
	
	In summary, phosphorus chemistry shows the greatest variability, particularly in \texttt{XODIAC-2025}, where different species dominate at different altitudes. In the deepest layers ($\sim$500~bar), \ce{PH3} is the most abundant species, followed by \ce{PH2}. 
	Near the top of the lower atmosphere ($\sim$1~bar), \ce{PH3} dominates, with \ce{P2} and \ce{PN} in second and third place, respectively. 
	At $\sim10^{-1}$~bar, \ce{P2} becomes most abundant, followed by \ce{PH3} and \ce{PN}. In the middle atmosphere ($\sim10^{-2}$~bar), \ce{PN} leads in abundance, followed by \ce{P2}, \ce{P}, \ce{PO}, and \ce{HOPO}. 
	In the upper middle atmosphere ($\sim10^{-4}$~bar), \ce{PN} remains dominant, with \ce{P} and \ce{PO} next, although both \ce{PO} and \ce{PH3} drop below a mixing ratio of $10^{-8}$. 
	In the uppermost layers ($\sim10^{-6}$ to $10^{-7}$~bar), atomic phosphorus (\ce{P}) is the principal species, while \ce{PN} and \ce{PO} are present but below the abundance threshold of $10^{-8}$. In contrast, \texttt{XODIAC-STAND2020} predicts a much simpler phosphorus chemistry, with \ce{PH3} and \ce{PH2} dominating in the lower and middle atmosphere, transitioning to \ce{PH2} in the upper atmosphere, where \ce{PH3} eventually falls below $10^{-8}$. The \texttt{XODIAC-SNCHO} network does not include phosphorus chemistry.

	\section{Conclusions} \label{sec:con}
	
	In this paper, we present \texttt{XODIAC}, a fast, parallelized, GPU-accelerated, and flexible 1D disequilibrium photochemical kinetics model with a built-in equilibrium chemistry solver (\texttt{NEXOCHEM}), updated thermochemical database,  and three integrated chemical reaction networks. This framework enables comparative assessments of planetary atmospheric chemistry within a single modeling environment. We applied \texttt{XODIAC} to the hot Jupiter HD~189733~b, yielding the following key findings:

	\begin{enumerate}
		\item The \texttt{XODIAC} model has been benchmarked against other widely used chemical kinetics models in the community, specifically \texttt{ARGO} \citep{Rimmer_2016} and \texttt{VULCAN} \citep{Tsai_2021}, using identical initial conditions and chemical networks for the hot Jupiter HD~189733~b. Our results show excellent agreement between models under these conditions.
		
		\item We compared different numerical formalisms, specifically the Lagrangian scheme used in \texttt{XODIAC} and \texttt{ARGO}, and the Eulerian scheme employed in \texttt{VULCAN}. When using the same chemical network and initial conditions, both approaches yield very similar outcomes, indicating that the choice of numerical scheme does not significantly affect the results.
		
		\item We introduced a state-of-the-art chemical network, \texttt{XODIAC-2025}, which builds upon the \texttt{STAND-2020} network \citep{Rimmer_2021} with substantial updates to phosphorus \citep{lee2024photochemical, DomagalGoldman2011, Silva_2025}, carbon \citep{willacy2022vertical}, sulfur \citep{tsai2024biogenic}, and sodium chemistry \citep{acharyya2024formation}. The final network includes 7,720 unique reactions involving 594 species, making it one of the most comprehensive \texttt{C-H-O-N-P-S-Metals} chemical networks developed to date for exoplanetary atmospheres. 
		
		\item We also introduce a newly compiled, state-of-the-art thermochemical database within the \texttt{XODIAC} framework, constructed by incorporating all species from the \texttt{STAND-2020} network \citep{Rimmer_2021}, along with 70 additional species. For three of these species (P($^2$D), \ce{P2O2}, and \ce{P2O}), NASA seven-coefficients have been calculated using quantum chemical theory.
		
		\item We implemented advanced parallel computing techniques to accelerate rate coefficient calculations in \texttt{XODIAC}, utilizing both CPU and GPU resources to efficiently handle the large chemical network. This optimization significantly reduces runtime and provides a scalable foundation for simulating even larger and more complex networks in the future. 
		
		\item The \texttt{XODIAC} model demonstrates a versatile treatment of photochemistry by allowing rate coefficients for photochemical reactions to be derived either directly from photodissociation/photoionization cross-section data, as in \texttt{ARGO}, or through an integrated approach that combines photodissociation/photoionization and photoabsorption cross-sections, as in \texttt{VULCAN}. This flexibility enables the model to accommodate different levels of data availability and scientific requirements, thereby broadening its applicability across a wide range of atmospheric studies. 
		
		\item Using \texttt{XODIAC}, we conducted a comparative study of three chemical networks (\texttt{XODIAC-2025}, \texttt{XODIAC-STAND2020}, and \texttt{XODIAC-SNCHO}) to investigate global carbon, sulfur, and phosphorus chemistry in the atmosphere of HD~189733~b and to identify their differences.
		
		\item Our analysis shows that \texttt{XODIAC-2025} captures key differences in hydrocarbon chemistry compared with the other networks, particularly in the relative abundances of \ce{C3H7} and \ce{CH3}. It also identifies the shortest \ce{CH4}-\ce{CO} and \ce{CH4}-\ce{CO2} pathways within the \texttt{XODIAC-2025} network.
		
		\item For sulfur chemistry, although minor differences appear in the predicted abundances of specific sulfur-bearing species, the overall trends remain consistent across all three networks, underscoring the robustness of \texttt{XODIAC} in modeling sulfur chemistry in HD~189733~b. We also identify the shortest pathway for forming \ce{SO} from \ce{H2S} in the upper atmosphere using the \texttt{XODIAC-2025} network.
		
		\item Finally, our results demonstrate that phosphorus chemistry exhibits strong altitude-dependent variability, with \texttt{XODIAC-2025} predicting complex transitions among species such as \ce{PH3}, \ce{P2}, and \ce{PN}, ultimately leading to atomic phosphorus in the upper atmosphere. In contrast, \texttt{XODIAC-STAND2020} produces a much simpler chemical profile dominated by \ce{PH3} and \ce{PH2}, while \texttt{XODIAC-SNCHO} omits phosphorus entirely. These differences highlight the sensitivity of atmospheric phosphorus predictions to network completeness and underscore the importance of employing updated chemical networks for accurate modeling.

	\end{enumerate}
	
	In summary, the \texttt{XODIAC} model, together with its state-of-the-art chemical network \texttt{XODIAC-2025} and newly compiled thermochemical database, enhanced with theoretically computed data from quantum chemical simulations, provides a powerful and versatile platform for advancing our understanding of exoplanetary atmospheric chemistry. This work establishes a solid foundation for future studies of increasingly complex and diverse planetary atmospheres.

	\section{acknowledgments}
	
	L.M. acknowledges funding support from the DAE through the NISER project RNI~4011. L.M. appreciates discussions with Sukit Ranjan regarding common convergence issues in solvers used in photochemical kinetics codes. Thanks are also due to Jeehyun Yang for valuable discussions on the use of the Reaction Mechanism Generator (RMG). L.M. further thanks Karen Willacy for insightful discussions related to hydrocarbon chemistry. P.G.\ and D.D.\ thank Dwaipayan Dubey for helpful discussions on developing an improved version of the Dijkstra algorithm for rate analysis, and they also thank Namrata Rani for assistance with Gaussian~16. P.G.\ also thanks S.~Maitrey for discussions on the \texttt{pylsodes} solver, which was developed for the \texttt{PEGASIS} astrochemical code and adapted for use in this work. S.M.\ expresses sincere gratitude to Mr.~Nishil Mehta for sharing his experience and expertise in disequilibrium chemistry in the early stages of the project. L.M.\ gratefully acknowledges support from Breakthrough Listen at the University of Oxford through a sub-award to NISER under Agreement R82799/CN013, provided as part of the global Breakthrough Listen collaboration funded by the Breakthrough Prize Foundation. We thank the anonymous referee for constructive comments that helped improve the manuscript.
	
	\vspace{5mm}
	
	\section{Data Availability}
	
	The \texttt{XODIAC-2025 C-H-O-N-P-S-Metals} network, together with the \texttt{XODIAC C-H-O-N-S} and \texttt{XODIAC C-H-O-N-S-Metals} networks in a unified format, will be made available upon reasonable request to the corresponding author.

	\facilities{JWST}
	\software{\texttt{VULCAN} \citep{Tsai_2021}, \texttt{ARGO} \citep{Rimmer_2016}, \texttt{Gaussian 16} \citep{g16}, \texttt{GaussView 6} \citep{dennington2016gaussview}, \texttt{RMG} \citep{johnson2022rmg}, \texttt{NumPy} \citep{harris2020array}, \texttt{matplotlib} \citep{Hunter:2007}, \texttt{pandas} \citep{mckinney2010data}}, \texttt{scipy} \citep{2020SciPy-NMeth}, \texttt{Numba} \citep{Kwan_numba}, \texttt{CuPy} \citep{Ryosuke_cupy_learningsys2017}
	
	\clearpage
	
	\twocolumngrid
	
	\section*{\textbf{APPENDIX A}}
	\phantomsection
	\label{app:apdx}

	\section*{Theoretical Calculation of Gibbs free energy}
	
	To evaluate the thermochemical properties of the studied species, a multi-step computational workflow was employed, involving molecular modeling, high-level quantum chemical calculations, and thermodynamic data processing.
	
	\subsection*{Molecular Structure Preparation}
	
	Initial geometries were built and visualized using \texttt{GaussView 6.0.16} \citep{dennington2016gaussview}, which provided an intuitive interface for constructing molecular structures and generating input files for quantum chemical simulations. The molecular symmetry and atom connectivity were carefully checked to ensure reliable optimization.
	
	\subsection*{Quantum Chemical Calculations}
	
	All electronic structure calculations were performed using \texttt{Gaussian~16 Rev.~A.03} \citep{g16}. Geometry optimizations and frequency calculations were carried out using the high-accuracy composite method CBS-QB3 for P$_2$O and P$_2$O$_2$, and M06-2X/aug-cc-pVTZ for the excited state P($^2$D), which provides a more reliable description of excited states compared to CBS-QB3 \citep{Zhao2008}. The spin multiplicity of each species was defined according to its ground-state electronic configuration, except for P($^2$D), ensuring the correct treatment of both closed- and open-shell systems.

	Frequency calculations were used to confirm that each optimized structure corresponds to a true local minimum (i.e., all real vibrational frequencies). These computations also provided zero-point vibrational energies (ZPVE), enthalpic corrections, and entropy contributions required for deriving temperature-dependent thermodynamic quantities.
	
	\subsection*{Thermodynamic Data Conversion}
	
	The thermodynamic data extracted from Gaussian output files were post-processed using the \texttt{Arkane} module of the \texttt{Reaction Mechanism Generator (RMG)} \citep{johnson2022rmg,liu2021rmg, dana2023_arkane} software suite (version 3.2.0). The resulting enthalpy ($H$), entropy ($S$), and heat capacity ($C_p$) values were fitted to the NASA 7-coefficient polynomial form:
	
	\begin{itemize}
		\item The first seven coefficients correspond to the low-temperature range (typically 10–1000 K),
		\item The second set of seven coefficients corresponds to the high-temperature range (typically 1000–3000 K).
	\end{itemize}
	
	This polynomial format enables seamless integration into kinetic models and facilitates simulations of temperature-dependent reaction networks.
	
	\subsection*{Validation of Computational Methodology}
	\begin{figure}[h!]
		
		\centering
		\includegraphics[width=\linewidth]{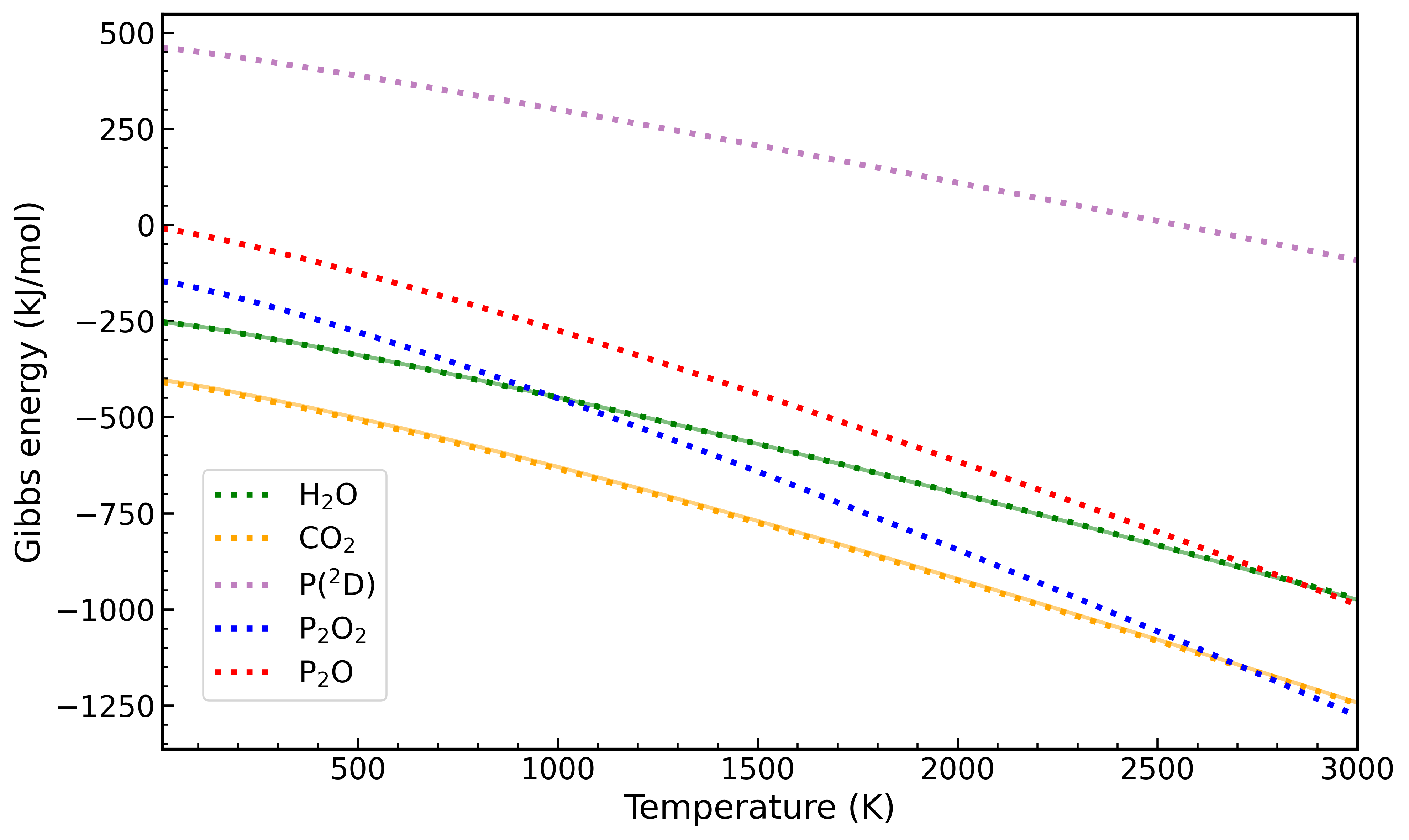}
		\caption{Comparison of Gibbs free energy ($G^\circ_T$, in kJ~mol$^{-1}$) as a function of temperature. Solid lines represent reference data from the Burcat thermochemical database for \ce{H2O} and \ce{CO2}, which are used to benchmark our computational method and results, whereas dotted lines correspond to all species computed in this work, including new species such as P($^2$D), P$_2$O$_2$, and P$_2$O, which are not present in the Burcat thermochemical database.}

		\label{g16-fig}
	\end{figure}
	
	\begin{figure}[h!]
		
		\centering
		\includegraphics[width=\linewidth]{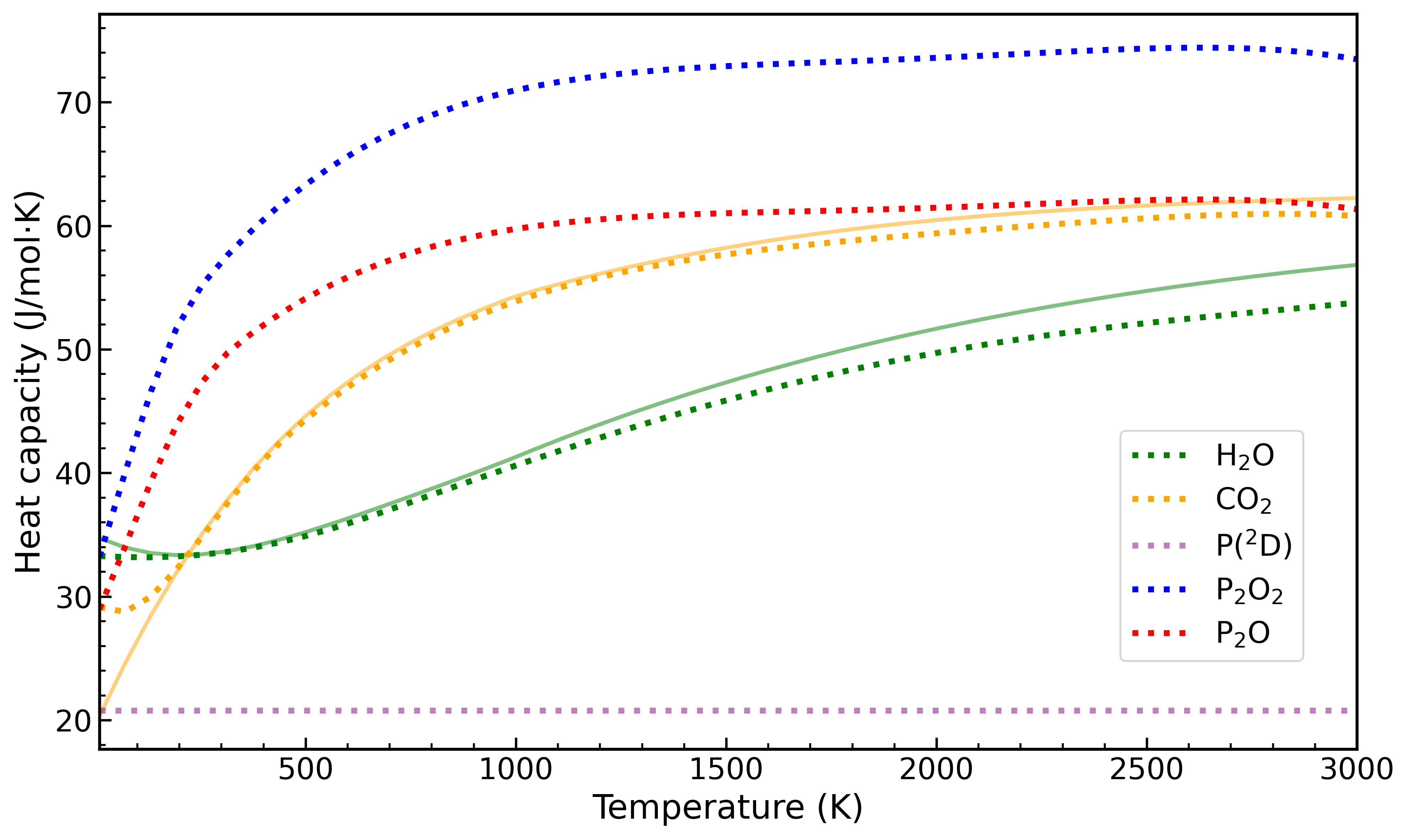}
		\caption{Comparison of Heat capacity ($C_p ^\circ$, in J/mol-K) as a function of temperature. Solid lines represent reference data from the Burcat thermochemical database for \ce{H2O} and \ce{CO2}, which are used to benchmark our computational method and results, whereas dotted lines correspond to all species computed in this work, including new species such as P($^2$D), P$_2$O$_2$, and P$_2$O, which are not present in the Burcat thermochemical database.}
		
		\label{cp-fig}
	\end{figure}
	
	To validate the thermochemical accuracy of our computational workflow, we benchmarked it against two well-characterized molecular species: water (\ce{H2O}) and carbon dioxide (\ce{CO2}). Their NASA 7-term polynomial coefficient data is taken from the Burcat thermochemical database~\cite{burcat_2005}.
	
	The dimensionless Gibbs free energy ($G_T^\circ/RT$) and heat capacity ($C_p^\circ/R$) were subsequently computed using the NASA polynomial coefficients over the temperature range of 10–3000~K. The corresponding expressions are given by:

	\[
	\frac{G^\circ_T}{RT} = a_1(1 - \ln T) - \frac{a_2 T}{2} - \frac{a_3 T^2}{6} - \frac{a_4 T^3}{12} - \frac{a_5 T^4}{20} + \frac{a_6}{T} - a_7
	\]
	
	\[
	\frac{C^\circ_p}{R} = a_1 + a_2 T + a_3 T^2 + a_4 T^3 + a_5 T^4
	\]
	where $G^\circ_T$ and $C^\circ_p$ are the standard Gibbs free energy and heat capacity at temperature $T$, $R$ is the ideal gas constant, and $a_1$ through $a_7$ are the NASA polynomial coefficients valid over specific temperature ranges.
	
	A comparison between the computed Gibbs free energy and heat capacity values and those from the Burcat database is shown in Figures~\ref{g16-fig} and~\ref{cp-fig}. The solid lines correspond to the reference data, while the dotted lines represent results obtained in this work. Numerical values of the NASA polynomial coefficients, both from Burcat and from our computations, are listed in Table~\ref{tab:nasa_coeffs} for benchmarking and for newly evaluated species. 
	\begin{table*}[h!]
		\centering
		\caption{NASA 7-coefficient polynomial parameters for selected molecules. First row corresponds to the low-temperature range; second row to the high-temperature range. The benchmark molecules (H\textsubscript{2}O and CO\textsubscript{2}) are shown with both original (\cite{burcat_2005}) and computed values.}
		\label{tab:nasa_coeffs}
		\normalsize  
		\begin{tabular}{lrrrrrrr}
			\toprule
			\textbf{Molecule} & \textbf{$a_1$} & \textbf{$a_2$} & \textbf{$a_3$} & \textbf{$a_4$} & \textbf{$a_5$} & \textbf{$a_6$} & \textbf{$a_7$} \\
			\midrule
			\multicolumn{8}{l}{\textbf{Benchmark Molecules}} \\
			\cmidrule{1-8}
			\ce{H_2O} (Burcat, low-T) & 4.1986 & -2.0364e-03 & 6.5203e-06 & -5.4879e-09 & 1.7720e-12 & -3.0294e+04 & -0.84901 \\
			\ce{H_2O} (Burcat, high-T) &  2.6770 & 2.9732e-03 & -7.7377e-07 & 9.4434e-11 & -4.2690e-15 & -2.9886e+04 & 6.8826 \\
			\ce{H_2O} (This work, low-T) & 4.00472 & -2.3916e-04 & 8.14071e-07 & 1.47085e-09 & -1.19281e-12 & -3.03697e+04 & -0.100539 \\
			\ce{H_2O} (This work, high-T) & 3.52201 & 1.05304e-03 & 6.39778e-07 & -3.85576e-10 & 5.48201e-14 & -3.0259e+04 & 2.3373 \\
			[0.2cm]
			\ce{CO_2} (Burcat, low-T) & 2.3568 & 8.9841e-03 & -7.1221e-06 & 2.4573e-09 & -1.4289e-13 & -4.8372e+04 & 9.9009 \\ 
			\ce{CO_2} (Burcat, high-T) & 4.6365 & 2.7415e-03 & -9.9590e-07 & 1.6039e-10 & -9.1620e-15 & -4.9025e+04 & -1.9349 \\
			\ce{CO_2} (This work, low-T) & 3.53569 & -3.56902e-03 & 4.01827e-05 & -7.42066e-08 & 4.49587e-11 & -4.90781e+04 & 5.39098 \\
			\ce{CO_2} (This work, high-T) & 2.78724 & 7.07379e-03 & -4.64997e-06 & 1.44009e-09 & -1.69498e-13 & -4.90661e+04 & 7.87818 \\
			[0.3cm]
			\multicolumn{8}{l}{\textbf{New Molecules}} \\
			\cmidrule{1-8}
			% P ($^2$D) (low-T) & 2.5000 & -1.5266e-15 & 6.3226e-18 & -7.5936e-21 & 2.6736e-24 & 5.5578e+04 & 4.6767 \\
			% P ($^2$D) (high-T) & 2.5000 & -1.0180e-13 & 8.3758e-17 & -2.9381e-20 & 3.7210e-24 & 5.5578e+04 & 4.6767 \\
			P ($^2$D) (low-T) & 2.5000 & -1.5266e-15 & 6.3226e-18 & -7.5936e-21 & 2.6736e-24 & 5.5578e+04 & 4.6767 \\
			P ($^2$D) (high-T) & 2.5000 & -1.0180e-13 & 8.3758e-17 & -2.9381e-20 & 3.7210e-24 & 5.5578e+04 & 4.6767 \\
			[0.2cm]
			% \ce{P2O2} (low-T) & 3.8696 & 1.2313e-02 & 2.9976e-05 & -2.1572e-07 & 3.0035e-10 & -1.7438e+04 & 8.9677 \\
			% \ce{P2O2} (high-T) & 5.1160 & 7.3694e-03 & -5.6388e-06 & 1.9346e-09 & -2.4535e-13 & -1.7564e+04 & 3.5937 \\
			\ce{P2O2} (low-T) & 3.87029 & 0.0121855 & 2.98192e-05 & -2.1085e-07 & 2.90231e-10 & -17417.5 & 8.96499 \\
			\ce{P2O2} (high-T) & 5.11029 & 0.00738589 & -5.65413e-06 & 1.94045e-09 & -2.46122e-13 & -17544 & 3.60157 \\
			[0.2cm]
			% \ce{P2O} (low-T) & 3.9272 & 5.9417e-03 & 2.3455e-05 & -1.0682e-07 & 1.1242e-10 & 1.0997e+04 & 9.9941 \\
			% \ce{P2O} (high-T) & 4.8722 & 4.2148e-03 & -3.3243e-06 & 1.1644e-09 & -1.4972e-13 & 1.0868e+04 & 5.5817 \\
			\ce{P2O} (low-T) & 3.39851 & 0.008556 & 2.65486e-05 & -1.34634e-07 & 1.48656e-10 & -968.182 & 7.99664 \\
			\ce{P2O} (high-T) & 4.58264 & 0.00574024 & -4.50982e-06 & 1.57536e-09 & -2.02135e-13 & -1120.92 & 2.55969 \\
			\bottomrule
		\end{tabular}
		\normalsize
	\end{table*}

	\begin{deluxetable*}{llll}
		\tablecaption{Compilation of species for which NASA-7 thermochemical coefficients are obtained from different sources, categorized
			by chemical family.}
		\tablehead{
			\multicolumn{1}{l}{\textbf{Species Type}} &
			\multicolumn{1}{l}{\textbf{Reaction Mechanism Generator}} &
			\multicolumn{1}{l}{\textbf{Burcat thermochemical database}} &
			\multicolumn{1}{l}{\citet{Venot2012}} \\
			\multicolumn{1}{l}{} &
			\multicolumn{1}{l}{\textbf{(RMG)}} &
			\multicolumn{1}{l}{\citep{burcat_2005}} &
			\multicolumn{1}{l}{}
		}

		\startdata
		P-bearing & 
		\parbox[t]{4cm}{--} &
		\parbox[t]{4cm}{CP, H$_2$POH, H$_3$PO$_4$, HOPO, HOPO$_2$, HPO, HPOH, P, P$_2$, P$_2$H, P$_2$H$_2$, P$_2$H$_4$, P$_2$O$_3$, P$_4$, P$_4$O$_6$, PH, PN, PO, PO$_2$, PO$_3$} &
		\parbox[t]{4cm}{--} \\
		% \hline
		\vspace{10pt}
		N-bearing & 
		\parbox[t]{4cm}{C$_3$N, C$_5$N, CH$_2$NH, CH$_3$NH, CH$_2$NH$_2$, CH$_3$C$_3$N, HC$_5$N} &
		\parbox[t]{4cm}{C$_2$N$_2$, C$_4$N$_2$, C$_6$N$_2$} &
		\parbox[t]{4cm}{N($^2$D)} \\
		\vspace{10pt}
		% \hline
		S-bearing & 
		\parbox[t]{4cm}{--} &
		\parbox[t]{4cm}{CH$_3$S$_2$CH$_3$, CH$_3$SCH$_2$, CH$_3$SCH$_3$, FeS} &
		\parbox[t]{4cm}{--} \\
		\vspace{10pt}
		% \hline
		Hydrocarbons / Others & 
		\parbox[t]{4cm}{C$_6$H$_2^+$, C$_6$H$_4^+$, C$_3$H$_4$, c-C$_3$H, c-C$_3$H$_2$, c-C$_3$H$_2^+$, c-C$_3$H$_3^+$, l-C$_3$H$_2$, l-C$_6$H$_6$, t-C$_3$H$_2$} &
		\parbox[t]{4cm}{1,2-C$_4$H$_6$, 1,3-C$_4$H$_6$, C$_3$, C$_3$H$_8$, C$_4$H$_5$, C$_4$H$_9$, C$_5$H$_3$, C$_5$H$_4$, C$_6$H, C$_6$H$_2$, C$_6$H$_3$, C$_6$H$_4$, C$_6$H$_5$, C$_6$H$_5^+$, C$_6$H$_6$, C$_6$H$_6^+$, C$_6$H$_7^+$, C$_8$H$_2$, l-C$_3$H, t-C$_3$H$_3^+$, CH$_3$Cl, HCCl} &
		\parbox[t]{4cm}{--} \\
		\enddata
	\end{deluxetable*}

	\clearpage

\bibliographystyle{aasjournal}
\bibliography{references}

\end{document}